\documentclass[10pt]{iopart}
\eqnobysec

\usepackage{iopams,graphicx,amssymb}

\def\eps{\varepsilon}
\def\Dm{\widetilde{\cal D}_{\mu}}
\def\const{{\rm const\,}}
\def\D{{\cal D}}

\def\S{{\cal S}}
\def\x{{\bf x}}
\def\p{{\bf p}}
\def\k{{\bf k}}
\def\q{{\bf q}}
\def\bfr{{\bf r}}

\def\n{{\bf n}}
\def\bfx{{\bf x}}
\def\bfv{{\bf v}}
\def\bfu{{\bf u}}

\begin{document}

\title[Critical behaviour of a fluid in a random shear flow]
{Critical behaviour of a fluid in a random shear flow:
Renormalization group analysis of a simplified model}


\author{N V Antonov and A A Ignatieva}

\address{Department of Theoretical Physics, St.~Petersburg University,
Uljanovskaja 1, St.~Petersburg, Petrodvorez, 198504 Russia}

\ead{nikolai.antonov@pobox.spbu.ru}

\begin{abstract}
Critical behaviour of a fluid (binary mixture or liquid crystal), subjected
to strongly anisotropic turbulent mixing, is studied by means of the field
theoretic renormalization group. As a simplified model, relaxational
stochastic dynamics of a non-conserved scalar order parameter, coupled to
a random velocity field with prescribed statistics, is considered. The
velocity is taken Gaussian, white in time, with correlation function of the
form $\propto \delta(t-t') /|{\bf k}_{\bot}|^{d+\xi}$, where ${\bf k}_{\bot}$
is the component of the wave vector, perpendicular to the distinguished
direction (``direction of the flow'') --- the $d$-dimensional generalization
of the ensemble introduced by Avellaneda and Majda [{\it Commun. Math. Phys.}
{\bf 131} 381] within the context of passive scalar advection. It is shown
that, depending on the relation between the exponent $\xi$ and the space
dimensionality $d$, the system exhibits various types of large-scale
self-similar behaviour, associated with different infrared attractive fixed
points of the renormalization-group equations. In addition to well known
asymptotic regimes (model A of equilibrium critical dynamics and passively
advected scalar with no self-interaction), existence of a new,
non-equilibrium and strongly anisotropic, type of critical behaviour
(universality class) is established, and the corresponding critical
dimensions are calculated to the second order of the double expansion
in $\xi$ and $\varepsilon=4-d$ (two-loop approximation). The most realistic
values of the model parameters (for example, $d=3$ and the Kolmogorov
exponent $\xi=4/3$) belong to this class. The scaling behaviour appears
anisotropic in the sense that the critical dimensions related to the
directions parallel and perpendicular to the flow are essentially different.
The results are in qualitative agreement with the results, obtained in
experiments and simulations of fluid systems subjected to various kinds of
regular and chaotic anisotropic flows.
\end{abstract}

\pacs{64.75.$+$g, 05.10.Cc, 64.60.Ht, 05.40$-$a}


\maketitle

\section{Introduction} \label{sec:Intro}

Various systems of very different physical nature
(ferromagnets and antiferromagnets, gas--vapour systems, binary liquid
mixtures and alloys) reveal interesting singular behaviour when undergoing
continuous (second-order) phase transition (that is, in the vicinity of
their critical points). Specific heat, susceptibility, spontaneous
magnetization {\it etc} exhibit singular self-similar (power-law)
behaviour, whose quantitative characteristics (critical dimensions and
scaling functions) depend only on few global characteristics of the system
(like symmetry or space dimensionality). This universality is related to
the existence in such systems of
a wide range of strongly coupled degrees of freedom: it produces a kind of
collective behaviour in which numerous irrelevant details of a specific
system are wiped away. This classical subject is exposed in the monographs
\cite{Fisher} and the literature cited therein.

Consistent qualitative and quantitative description of the equilibrium
critical behaviour was achieved within the framework of the
renormalization group (RG). In the RG approach, possible types of critical
regimes (universality classes) are associated with infrared (IR) attractive
fixed points of renormalizable field theoretic models. Most typical phase
transitions belong to the universality class of the
$O(N)$-symmetric $\varphi^{4}$ model of an $N$-component scalar order
parameter (Landau--Ginzburg Hamiltonian). Universal characteristics of the
critical behaviour depend only on $N$ and the space dimensionality $d$ and
can be calculated in the form of the expansion in $\eps=4-d$ or within other
systematic perturbation schemes; see the monographs \cite{Zinn,Book3} and
the literature cited therein.

Dynamical critical behaviour (critical singularities of relaxation and
correlation times, various kinetic and transport coefficients {\it etc})
appears richer, less universal and is comparatively less understood.
Different nature of the order parameter (conserved or non-conserved),
inclusion of ``secondary'' slow modes (densities of entropy or energy) and
interaction with hydrodynamical degrees of freedom produce different types
of critical dynamics for the same static model \cite{Book3,HH,MF}.
The reliable sets of second-order (two-loop) results were only recently
fixed, and many important questions remain open; see the recent review
paper \cite{MF} for discussion and bibliography.

It has long been realized that the behaviour of a real system near its
critical point is extremely sensitive to external disturbances, geometry
of the experimental setup, gravity, presence of impurities and so on.
``Ideal'' equilibrium critical behaviour of an infinite system can be
obscured by limited accuracy of measuring the temperature, finite-size
effects, finite time of evolution (ageing) and so on. What is more, some
disturbances (randomly distributed impurities in magnets and turbulent
mixing of fluid systems) can produce completely new types of critical
behaviour with rich and rather exotic properties (e.g., expansion in
$\sqrt{\eps}$ rather than in $\eps$); see \cite{quench,Satten,Ronis}.
Over the past three decades, considerable attention has been attracted
by the effects of various kinds of imposed flows (laminar shear flows,
turbulent stirring and other types of deterministic or chaotic flows)
on the behaviour of critical fluids, e.g., of binary liquid mixtures near
the consolution point; see the papers \cite{Satten}--\cite{Mixing} and
references therein. This problem is closely related to another interesting
issue: the effects of imposed flows on the dynamics of phase ordering ---
the growth of order through domain coarsening (spinodal decomposition),
when a system is quenched from its high-temperature homogeneous phase
into the low-temperature multi-phase coexistence region; see e.g.
\cite{Shear2}--\cite{Chaotic} and the literature cited therein.

Although very different, many of such systems exhibit a common interesting
feature: existence of new non-equilibrium stationary states with (arguably)
self-similar statistical properties and new sets of scaling exponents.
Emergence of nonequilibrium steady states appears rather a generic and
robust phenomenon, being observed in experiments, simulations and
analytical treatments of critical liquids with passive and active order
parameters subjected to laminar or turbulent flows, various kinds of regular
and chaotic synthetic velocity ensembles, cellular or shear flows, and so on
\cite{Satten}--\cite{Mixing}. In the presence of a distinguished direction,
scaling behaviour of such systems appears strongly anisotropic, with
different critical dimensions corresponding to different spatial directions
\cite{Shear2}--\cite{Shear1}.

The aim of the theory is to establish existence of such regimes on the
basis of microscopic dynamic models, to classify corresponding universality
classes, to calculate their scaling dimensions within consistent
approximations or regular perturbation schemes, to investigate their
universality, dependence on the model parameters, and so on.
In this paper, we will focus on the {\it anisotropic} turbulent mixing
of critical fluids, because most real flows are strongly anisotropic,
this anisotropy persists in the asymptotic critical regime and leads
to new interesting effects.

The full-scale model of a critical fluid subjected to a strongly anisotropic
turbulent stirring must deal with a conserved (binary mixtures) or
non-conserved (liquid crystals) order parameter with mutual coupling
with the velocity field, governed by nonlinear dynamic equations (e.g.
stochastic Navier--Stokes equation with an external random stirring
force), and the anisotropy is introduced by the initial and/or boundary
conditions. However, even for the equilibrium and isotropic case (model H
in the traditional classification introduced in \cite{HH}), the consistent
RG analysis of such problem appears a most difficult task, and has only
recently been completed (see discussion and references in \cite{MF} and
sections 5.23--5.25 in book \cite{Book3}), while theoretical description
of fully developed turbulence on the basis of dynamic equations remains,
in many respects, essentially an open problem.

In the present paper, we apply the field theoretic RG to a
simplified ``minimal'' model of a stirred critical fluid, which
nevertheless appears rather nontrivial and captures the main
property of the problem: existence of a new, non-equilibrium and
strongly anisotropic, universality class of scaling behaviour.
Namely, we consider a purely relaxational dynamics of a
non-conserved passive scalar order parameter (model A in
terminology of \cite{HH}) coupled to the random velocity field
with prescribed Gaussian statistics.

Recently, the models involving passive (no feedback on the velocity) linear
(no self-interaction) scalar fields advected by such ``synthetic'' velocity
ensembles attracted enormous attention among the ``turbulent community''
because of the insight they offer into the origin of intermittency and
anomalous scaling in the real fluid turbulence; see the review paper
\cite{FGV} and references therein. In spite of their relative simplicity,
such models reproduce many of the anomalous features of genuine turbulent
heat or mass transport observed in experiments. Most popular is the
Kazantsev--Kraichnan ensemble with the velocity correlation function of
the form $\langle vv\rangle \propto \delta(t-t') \, k^{-d-\xi}$. Vanishing
of the correlation time is necessary to ensure Galilean symmetry of the
problem, while a power-law dependence on the wave number $k$ mimics real
self-similar properties of fully developed turbulence. For a {\it conserved}
order parameter, it can be shown that the nonlinearity in the Navier-Stokes
equation and a finite correlation time are indeed IR irrelevant (in the
sense of Wilson) in the analysis of critical behaviour; see also the
discussion in \cite{Ronis}.

In the RG approach to the Kraichnan model and its descendants, reviewed in
\cite{JphysA}, the exponent $\xi$ plays the role of a formal RG expansion
parameter, analogous in this respect to the conventional $\eps=4-d$.

Synthetic ensembles also allow one to easily introduce anisotropy,
compressibility {\it etc}, and to study their effects on the
behaviour of the scalar field. In this paper we employ the
$d$-dimensional generalization of a strongly anisotropic ensemble
introduced in \cite{AM} in connection with the passive linear
problem: the velocity field is oriented along a chosen direction
$\n$ and its correlation function depends only on the coordinates
perpendicular to $\n$; see also Refs.~\cite{Glimm,AM2}.

In experiments, critical fluids with a non-conserved order parameter
can be realized in twisted nematic liquid crystals; see the discussion
and references in \cite{Shear3,Shear1}.
In a wider context, such model can be viewed as a model system for studying
generic nonequilibrium phase transitions. Recently, significant progress
has been achieved in classifying IR scaling behaviour of such phenomena,
including driven diffusive systems, diffusion-limited reactions, growth,
ageing and percolation processes, and so on; see e.g. Refs.
\cite{driven}--\cite{obliq} and references therein. Being analytically
tractable, our model can serve as a possible testing ground in studying
such scaling regimes and their universality within controlled approximations
or a regular perturbative scheme.

Earlier, the field theoretic RG was applied to the stirred critical fluid
in a number of studies. New types of critical behaviour were identified for
model B in a Gaussian velocity ensemble \cite{Satten,Ronis} and generalized
model A with inclusion of a large-scale stirring force and the velocity
field governed by the stochastic Navier--Stokes equation \cite{IJMP,Mixing},
but only purely isotropic situations were considered. The RG ideas were
also applied to the problem of phase separation and domain growth below
the critical temperature; see e.g. \cite{RGDom,Adv} and references therein.
By contrast with critical phenomena, the RG approach to such problems
suffers from the lack of an (obvious) small parameter (like $\eps=4-d$
or $\xi$ in our case) and should involve numerical (Monte Carlo) simulations
\cite{RGDom} or additional phenomenological hypotheses \cite{Adv}.

The paper is organized as follows. In section~\ref{sec:QFT} we give detailed
description of the model, present its field theoretic formulation and the
corresponding diagrammatic technique. In section~\ref{sec:Reno} we analyze
canonical dimensions and ultraviolet (UV) divergences of the model. We show
that, after an appropriate extension, the model becomes multiplicatively
renormalizable and present the corresponding renormalized action functional.
We also show that, for the extended model, independent canonical dimensions
should be introduced for the directions parallel and perpendicular to the
flow. In section~\ref{sec:RGE} we derive the differential RG equations,
introduce the RG functions ($\beta$ functions and anomalous dimensions
$\gamma$) and give the corresponding one-loop and two-loop expressions for
the case of an $N$-component order parameter. In section~\ref{sec:FPS} we
analyze possible scaling regimes of the model, associated with the fixed
points of the RG equations, and identify their ranges of stability in the
$\eps$--$\xi$ plane. Three fixed points correspond to known regimes: free
(Gaussian) regime, linear passive scalar and equilibrium model A. The fourth
fixed point corresponds to a new, non-equilibrium and strongly anisotropic,
universality class. For the most realistic values of the model parameters
(namely, $d=3$ and $0<\xi<2$; see section~\ref{sec:QFT}) it is the latter
point that is IR attractive and governs the large-scale, long-time behaviour
of the system. The corresponding critical dimensions depend on $\eps$ and
$\xi$ and can be systematically calculated as double series in those
parameters; the explicit second-order results are presented in
section~\ref{sec:DimeNS}. The scaling regime appears strongly anisotropic
in the sense that the critical dimensions
related to the directions parallel and perpendicular to the flow are
different. Section~\ref{sec:Conc} is reserved for discussion, comparison to
the existing experimental and theoretical results and the conclusions.

In appendix A we explore consequences of the Galilean symmetry for the
renormalization of our model. The main points concerning the calculation
of the renormalization constants and RG functions are briefly discussed
in appendix B.

\section{The model. Field theoretic formulation} \label{sec:QFT}

Relaxational dynamics of a non-conserved scalar order parameter
$\varphi(x)$ with $x \equiv \{t,\bfx\}$ is described by a
stochastic differential equation
\begin{eqnarray}
\sigma_{0}\partial_{t} \varphi(x) = - \delta {\cal H}(\varphi) /
\delta \varphi(x) + f(x),
\label{eq1}
\end{eqnarray}
where $\partial_{t} = \partial/ \partial {t}$, $\sigma_{0}
=1/\Gamma_{0}$ is the reciprocal of the (constant) kinetic
coefficient $\Gamma_{0}>0$ and $f(x)$ is a Gaussian random noise
with zero mean and the pair correlation function
\begin{eqnarray}
D_{f}\,(x-x') \equiv \langle f(x)f(x') \rangle = 2 \sigma_{0}
\delta(t-t')\, \delta^{(d)}(\x-\x') ,
\label{forceD}
\end{eqnarray}
$d$ being the dimensionality of the $\bfx$ space. Near the critical point,
the Hamiltonian ${\cal H}(\varphi)$ is taken in the Landau--Ginzburg form
\begin{eqnarray}
{\cal H}(\varphi) =  \int d{\x} \left\{- \frac{1}{2}\,
 \varphi(\x)\partial^{2} \varphi(\x) + \frac{\tau_{0}}{2}\,
\varphi^{2}(\x) + \frac{\lambda_{0}}{4!}\, \varphi^{4}(\x)
\right\},
\label{LG}
\end{eqnarray}
where $\partial_{i}= \partial/\partial x_{i}$ is the spatial derivative,
$\partial^{2}=\partial_{i}\partial_{i}$ is the Laplacian,
$\tau_{0} \propto (T-T_{c})$ measures deviation from the critical
temperature and $\lambda_{0}>0$ is the coupling constant;
after the functional differentiation in (\ref{eq1}) one has to replace
$\varphi(\bfx) \to \varphi(x)$. The model (\ref{eq1})--(\ref{LG}) is
referred to as model A \cite{HH}; its critical behaviour is very well
understood \cite{Zinn,Book3,HH,MF}.

Coupling with the velocity field $v_{i}(x)$ is introduced by the replacement
\begin{eqnarray}
\partial_{t} \to \nabla_{t} = \partial_{t} + v_{i} \partial_{i},
\label{nabla}
\end{eqnarray}
where $\nabla_{t}$ is the Lagrangian (Galilean covariant) derivative.

Let $\n$ be a unit constant vector that determines distinguished direction
(``direction of the flow''). Then any vector can be decomposed into the
components perpendicular and parallel to the flow, for example,
$\x = \x_{\bot} + \n x_{\parallel}$ with $\x_{\bot} \cdot \n =0$.
The velocity field will be taken in the form
\begin{eqnarray}
\bfv = {\bf u} + \n v(t, \x_{\bot}),
\label{vello}
\end{eqnarray}
where ${\bf u}$ is a constant
vector parallel to $\n$ and $v(t, \x_{\bot})$ is a scalar function
independent of $x_{\parallel}$. Then the incompressibility condition
is automatically satisfied:
\begin{eqnarray}
\partial_{i} v_{i} = \partial_{\parallel} v(t, \x_{\bot}) = 0.
\label{inko}
\end{eqnarray}
From now on, we set ${\bf u}=0$ (the general case ${\bf u} \ne 0$ leads to no
serious alterations in the RG analysis and will be briefly discussed in the
end of section~\ref{sec:Conc}). For $v(t, \x_{\bot})$ we assume a Gaussian
distribution with zero mean and the pair correlation function of the form:
\begin{eqnarray}
\langle v(t,\x_{\bot}) v(t',\x_{\bot}') \rangle = \delta(t-t') \int
\frac{d \k}{(2\pi)^{d}} \, \exp \left\{ {\rm i} \k\cdot (\x-\x')
\right\}
D_{v} (k)= \nonumber \\
= \delta(t-t') \int \frac{d \k_{\bot}}{(2\pi)^{d-1}} \, \exp
\left\{ {\rm i} \k_{\bot}\cdot (\x_{\bot}-\x'_{\bot}) \right\}
\widetilde D_{v} (k_{\bot}) , \quad  k_{\bot}=|\k_{\bot}|
\label{veloc1}
\end{eqnarray}
with the scalar coefficient functions of the form
\begin{eqnarray}
D_{v} (k)= 2\pi \delta(k_{\parallel}) \, \widetilde D_{v} (k_{\bot}) ,
\quad \widetilde D_{v} (k_{\bot}) = D_{0}\, k_{\bot}^{-d+1-\xi}.
\label{veloc2}
\end{eqnarray}
Here $D_{0}>0$ is a constant amplitude factor and $\xi$ an arbitrary
exponent, which (along with the conventional $\eps=4-d$) will play the part
of a formal RG expansion parameter. The IR regularization in (\ref{veloc1})
is provided by the cutoff $k_{\bot}>m$ (by dimension, $\tau_{0} \propto
m^{2}$). [Precise form of the IR regularization is inessential;
sharp cutoff is the most convenient choice from the calculational viewpoints.
Another possibility is to replace $k_{\bot}\to\sqrt{k^{2}_{\bot}+m^{2}}$
in (\ref{veloc2}).] The natural interval for the
exponent is $0< \xi <2 $, when the so-called ``effective eddy diffusivity''
\begin{eqnarray}
{\cal V}(\bfr_{\bot}) = \int \frac{d \k_{\bot}}{(2\pi)^{d-1}} \, \left\{ 1-
\exp \left( {\rm i} \k_{\bot}\cdot \bfr_{\bot}  \right) \right\} \,
\widetilde D_{v} (k_{\bot})
\label{effect}
\end{eqnarray}
has a finite limit for $m\to0$; it includes the most realistic Kolmogorov
value $\xi=4/3$. The exponent $\xi$ can also be viewed as a kind of
H\"{o}lder exponent, which measures ``roughness'' of the velocity field
\cite{FGV}; the ``Batchelor limit'' $\xi\to2$ corresponds to smooth velocity.

In order to ensure multiplicative renormalizability of the model, it is
necessary to split the Laplacian in (\ref{LG}) into the parallel and
perpendicular parts $\partial^{2} \to \partial^{2}_{\bot}
+ u_{0} \partial^{2}_{\parallel}$ by introducing a new parameter
$u_{0}>0$ (in the anisotropic case, these two terms will be renormalized
in a different way). Thus equation (\ref{eq1}) becomes
\begin{eqnarray}
\sigma_{0}\nabla_{t} \varphi(x) =  \partial^{2}_{\bot} \varphi(x) +
u_{0} \partial^{2}_{\parallel} \varphi(x) -\tau_{0} \varphi(x) -
\lambda_{0} \varphi^{3}(x) /6 +f(x) ;
\label{eq2}
\end{eqnarray}
this completes formulation of the model.

Interpretation of the splitting of the Laplacian term in (\ref{eq2}) can be
twofold. On the one hand, the fluctuation models of the type (\ref{eq1}) and
(\ref{LG}) are phenomenological and, by construction, they must require all
the IR relevant terms allowed by symmetry. The fact that the splitting is
required by the renormalization procedure means that it is not forbidden by
dimensionality or symmetry considerations and, therefore, it is natural to
include the general value $u_{0}\ne1$ to the model from the very beginning.
On the other hand, one can insist on studying the original model with
$u_{0} =1$ and $SO(d)$ covariant Laplacian term, although that symmetry is
broken to $SO(d-1)$ by the interaction with the anisotropic velocity
ensemble. Then the extension of the model to the case $u_{0}\ne1$ can be
viewed as a purely technical trick which is only needed to ensure the
multiplicative renormalizability and to derive the RG equations. The latter
should be then solved with the special initial data corresponding to
$u_{0} =1$ (in renormalized variables this anyway will correspond to
general initial data with $u\ne 1$). Since the IR attractive fixed point
of the RG equations is unique for any given choice of the parameters
$\eps$ and $\xi$ (see section~\ref{sec:FPS}), the resulting IR behaviour
will be the same as for the case of the extended model with general
$u_{0}\ne1$.

According to the general theorem \cite{MSR} (see also the monographs
\cite{Zinn,Book3}), our stochastic problem is equivalent to the field
theoretic model of the extended set of fields $\Phi = \{ \varphi', \varphi,
{\bf v} \}$ with action functional
\begin{eqnarray}
\S(\Phi) = \sigma_{0} (\varphi')^{2} + \varphi' \left[
- \sigma_{0} \nabla_{t} \varphi + \partial^{2}_{\bot} \varphi +
u_{0} \partial^{2}_{\parallel} \varphi -\tau_{0} \varphi -
\lambda_{0} \varphi^{3} /6 \right] + \S_{v} (\bfv) .
\label{action}
\end{eqnarray}
The first few terms represent the De Dominicis--Janssen action functional
for the stochastic problem (\ref{eq1}), (\ref{forceD}) at fixed ${\bf v}$;
it involves auxiliary scalar response field $\varphi'(x)$.  All the required
integrations over $x=\{t,{\bf x}\}$ and
summations over the vector indices are implied, for example,
\[ \varphi' \partial^{2}_{\bot} \varphi = \int dt \int d\x \,
\varphi'(x) \partial^{2}_{\bot} \varphi(x).  \]
It is worth noting that, owing to transversality of the velocity field
(\ref{inko}), the derivative in the coupling term in (\ref{action}) can
also be moved onto the field $\varphi'$ using integration by parts:
\begin{eqnarray}
\varphi' (v_{i} \partial_{i}) \varphi  = \int dt \int d\x \,
\varphi'(x) v(t,\x_{\bot}) \partial_{\parallel} \varphi(x) =
\nonumber \\
= - \int dt \int d\x\, (\partial_{\parallel} \varphi'(x)) v(t,\x_{\bot})
\varphi(x).
\label{Villon}
\end{eqnarray}

The last term in (\ref{action}) corresponds to the Gaussian averaging over
${\bf v}$ with correlator (\ref{veloc1}) and has the form
\begin{eqnarray}
\S_{v} (\bfv) = \frac{1}{2}\,
\int dt \int d\x_{\bot} d\x_{\bot}' v(t,\x_{\bot})
\widetilde D^{-1}_{v} (\x_{\bot}-\x'_{\bot}) v(t,\x_{\bot}'),
\label{Sv}
\end{eqnarray}
where
\begin{eqnarray}
\widetilde D^{-1}_{v} (\bfr_{\bot}) \propto D_{0}^{-1} \, r_{\bot}^{2(1-d)-\xi}
\label{Dv}
\end{eqnarray}
is the kernel of the inverse linear operation $D^{-1}_{v}$ for the
correlation function $D_{v}$ in (\ref{veloc2}).

This formulation means that statistical averages of random quantities
in the original stochastic problem coincide with the Green functions of the
field theoretic model with action (\ref{action}), given by functional
averages with the weight $\exp {\cal S}(\Phi)$ (see equation (\ref{GW})
in the appendix A). This allows one to apply
the field theoretic renormalization theory and renormalization group to
our stochastic problem. The model (\ref{action}) corresponds to a standard
Feynman diagrammatic technique with three bare propagators (lines in the
diagrams): $\langle \bfv\bfv \rangle_{0}$, given by (\ref{veloc1}),
(\ref{veloc2}), and the propagators of the scalar fields
(in the frequency--momentum and time--momentum representations):
\begin{eqnarray}
\langle \varphi \varphi' \rangle_{0} = \langle \varphi' \varphi
\rangle_{0}^{*} =
\nonumber \\
= \left\{-{\rm i} \sigma_{0} \omega+ \epsilon(\k)
\right\}^{-1} \ \leftrightarrow\ \theta(t-t')\, \sigma_{0}^{-1} \,
\exp{ \{- \epsilon(\k) (t-t')/\sigma_{0}\} },
\label{lines3}
\end{eqnarray}
where $\epsilon(\k) = \k_{\bot}^{2}+u_{0}\k_{\parallel}^{2}+\tau_{0}$ and
$\theta(\dots)$ is the Heaviside step function, and
\begin{eqnarray}
\langle \varphi \varphi \rangle_{0} =  2 \sigma_{0} \,
\left\{ \omega^{2}\sigma_{0}^{2} + \epsilon^{2}(\k) \right\}^{-1}
\quad\leftrightarrow\quad
\frac{1}{\epsilon(\k) } \, \exp{ \{- \epsilon(\k) |t-t'|/\sigma_{0}\} };
\label{lines2}
\end{eqnarray}
the propagator $\langle \varphi' \varphi' \rangle_{0}$ vanishes identically
for any field theory of the type (\ref{action}). The model also
involves two types of vertices corresponding to the interaction terms
$\varphi' \varphi^{3}$ and $\varphi'(v \partial_{\parallel}) \varphi$.
The corresponding coupling constants (``charges'')
$g_{0}$ and $w_{0}$ defined are introduced by the relations
\begin{eqnarray}
\lambda_{0} = u_{0}^{1/2} g_{0}, \quad  D_{0} = w_{0} u_{0} /\sigma_{0},
\label{D0}
\end{eqnarray}
so that by dimension $g_{0} \sim \ell^{-\eps}$ and $w_{0} \sim \ell^{-\xi}$,
where is $\ell$ has the order of the smallest length scale of our problem.
More precisely, these two lengths are rather different: the coupling
$g_{0}$ in the Landau--Ginzburg model is conventionally
related to the molecular length,
while $w_{0}$ corresponds to the Kolmogorov (dissipation) scale of
turbulence. However, in the following we will be interested in the
behaviour of the correlation functions at distances much larger than the
both these lengths, which allows us not to distinguish them. Thus we can
write
\begin{eqnarray}
g_{0} \sim \Lambda^{\eps}, \quad  w_{0} \sim \Lambda^{\xi},
\label{g0}
\end{eqnarray}
where $\Lambda$ sets the characteristic UV momentum scale.

By rescaling the fields, the coupling constant $w_{0}$ can be placed in
front of the interaction term $\varphi'(v \partial) \varphi$ in the action
(\ref{action}), which is more familiar for the field theory. We do not
do it, however, in order not to spoil the natural form of the covariant
derivative, and assign the factor $w_{0}$ to the propagator
$\langle \bfv\bfv \rangle_{0}$.

\section{Canonical dimensions and renormalization} \label{sec:Reno}

It is well known that the analysis of UV divergences is based on the analysis
of canonical dimensions (``power counting''); see e.g. \cite{Zinn,Book3}.
General dynamic models of the type (\ref{action}), in contrast to static
models (like e.g. (\ref{LG})), have two scales: canonical dimension of some
quantity $F$ (a field or a parameter in the action functional) is completely
characterized by two numbers, the frequency dimension $d_{F}^{\omega}$ and
the momentum dimension $d_{F}^{k}$. They are determined such that
$[F] \sim [T]^{-d_{F}^{\omega}} [L]^{-d_{F}^{k}}$, where $L$ is the length
scale and $T$ is the time scale; see e.g. Chap.~5 in book \cite{Book3}.
Our strongly anisotropic model, however, has two independent momentum scales,
related to the directions perpendicular and parallel to the vector $\n$, and
a more detailed specification of the canonical dimensions is necessary.
Namely, one has to introduce two independent momentum canonical dimensions
$d_{F}^{\bot}$ and $d_{F}^{\parallel}$ so that
\[ [F] \sim [T]^{-d_{F}^{\omega}}  [L_{\bot}]^{-d_{F}^{\bot}}
[L_{\parallel}]^{-d_{F}^{\parallel}}, \]
where $L_{\bot}$ and $L_{\parallel}$ are (independent) length scales in the
corresponding subspaces. The dimensions are found from the obvious
normalization conditions $d_{k_{\bot}}^{\bot}= -d_{\bf x_{\bot}}^{\bot}=1$,
$d_{k_{\bot}}^{\parallel}=-d_{\bf x_{\bot}}^{\parallel}=0$,
$d_{k_{\bot}}^{\omega} = d_{k_{\parallel}}^{\omega}=0$,
$d_{\omega }^{\omega }=-d_t^{\omega }=1$, and so on, and from the
requirement that each term of the action functional (\ref{action})
be dimensionless (with respect to all the three independent dimensions
separately). The original momentum dimension can be found from the
relation $d_{F}^{k} = d_{F}^{\bot}+ d_{F}^{\parallel}$.
Then, based on $d_{F}^{k}$ and $d_{F}^{\omega}$, one can introduce the
total canonical dimension $d_{F}=d_{F}^{k}+2d_{F}^{\omega}  =
d_{F}^{\bot} + d_{F}^{\parallel} +2d_{F}^{\omega}$ (in the free theory,
$\partial_{t}\propto\partial^{2}_{\bot} \propto \partial^{2}_{\parallel}$),
which plays in the theory of renormalization of dynamic models the same
role as the conventional (momentum) dimension does in static problems;
cf. Chap.~5 in book \cite{Book3}.

The full set of independent
canonical dimensions is needed, in particular, to identify the completely
dimensionless parameters, which only can appear as arguments in the
renormalization constants and RG functions. Of course, existence of several
independent spatial scales is not too exotic; it was encountered in a
number of models: ferroelectrics \cite{ferro} (see also section~1.17 of book
\cite{Book3}), continuous models of self-organized criticality \cite{Tadic},
anisotropic versions of the Kardar--Parisi--Zhang model \cite{KimKim},
$m$-axial Lifshits points \cite{maxial} and growing surfaces, driven by
obliquely incident particle beams \cite{obliq}.

The canonical dimensions for the model (\ref{action}) are summarized in
table~\ref{table1}, including renormalized parameters, which will be
introduced later on. From table~\ref{table1} or, equivalently, from
the relations (\ref{g0}), it follows that the model is logarithmic (the
coupling constants $g_{0}$ and $w_{0}$  are simultaneously dimensionless)
at $d=4$ and $\xi=0$, so that the UV divergences in the correlation
functions manifest themselves as poles in $\eps \equiv 4-d$, $\xi$ and
their linear combinations or, in general, as singularities at $\eps$ and
$\xi \to 0$.

\begin{table}
\caption{Canonical dimensions of the fields and parameters in the
model (\protect\ref{action})} \label{table1}
\begin{tabular}{cccccccccccc}
\br
$F$ & $\varphi $ & $\varphi'$ & $ {\bf v} $ & $\sigma ,\sigma_{0}$
& $u, u_{0}$ & $m,\mu, \Lambda $ & $g_{0}$ & $w_{0}$ & $g,w$ \\
\br
$d_{F}^{\omega}$ & 0 & 1 & 1 & $-1$ & 0 & 0 & 0 & 0& 0\\
\mr
$d_{F}^{\bot}$ & $(d-3)/2$ & $(d-3)/2$ & 0 & 2 & 2& 1 & $4-d$ & $\xi$ & 0 \\
\mr
$d_{F}^{\parallel}$ & $1/2$ & $1/2$ & $-1$ & 0 & $-2$ & 0 & 0 & 0 & 0 \\
\mr
$d_{F}^{k}=d_{F}^{\bot}+d_{F}^{\parallel}$ & $d/2-1$ & $d/2-1$ & $-1$ &
2 & 0 & 1 & $4-d$ & $\xi$ & 0 \\
\mr
$d_{F}=2 d_{F}^{\omega}+d_{F}^{k}$ & $d/2-1$ & $d/2+1$ & 1 & 0 & 0& 1 &
$4-d$ & $\xi$ & 0 \\
\br
\end{tabular}
\end{table}

The total canonical dimension of an arbitrary 1-irreducible Green function
$\Gamma = \langle\Phi \cdots \Phi \rangle _{\rm 1-ir}$ is given by the
relation
\begin{equation}
d_{\Gamma }= d+2- \sum_{\Phi} N_{\Phi }d_{\Phi}, \quad
\sum_{\Phi} N_{\Phi} d_{\Phi} = N_{\varphi'} d_{\varphi'} +
N_{\varphi} d_{\varphi} + N_{\bfv} d_{\bfv}.
\label{dGamma}
\end{equation}
Here $N_{\Phi}=\{N_{\varphi},\,N_{\varphi'},\,N_{\bf v}\}$ are the numbers of
corresponding fields entering into the function $\Gamma$, and the summation
over all types of the fields in (\ref{dGamma}) and analogous formulas below
is always implied.

The total dimension $d_{\Gamma}$ in logarithmic theory (that is, at
$\eps=\xi=0$) is the formal index of the UV divergence
$\delta_{\Gamma}=d_{\Gamma}|_{\eps=\xi=0}$.
Superficial UV divergences, whose removal requires counterterms, can be
present only in those functions $\Gamma$ for which $\delta_{\Gamma}$ is
a non-negative integer. The counterterms are local, that is, in the
frequency-momentum representation the counterterm to a given function
$\Gamma$ is a polynomial in $\omega$, $\k_{\bot}$ and $\k_{\parallel}$.
Since the parameters $u_{0}$ and $\sigma_{0}$ are dimensionless with
respect to the total dimension $d_{F}$, the index $\delta_{\Gamma}$
gives the degree of that polynomial (with the assumption that
$\omega \sim k_{\bot}^{2} \sim k_{\parallel}^{2}$); detailed structure
of the counterterms and their dependence on $\sigma_{0}$ and $u_{0}$
is directly found from the corresponding partial dimensions
$d_{\Gamma}^{\omega}$ and $d_{\Gamma}^{\bot,\parallel}$.
From table~\ref{table1} and (\ref{dGamma}) we find
\begin{equation}
\delta_{\Gamma}= 6 - 3 N_{\varphi'} - N_{\varphi} -N_{v}.
\label{IndeX}
\end{equation}

Dimensional considerations should be augmented by the observation that all
the 1-irreducible functions without the field $\varphi'$ (in particular,
all functions involving only velocity fields) contain closed circuits of
retarded propagators $\langle \varphi' \varphi  \rangle_{0}$, vanish and do
not require counterterms; see e.g. \cite{Book3}. The action (\ref{action})
is even with respect to the reflection $\varphi' \to - \varphi'$,
$\varphi \to - \varphi$, so that all correlation functions with odd total
number of the fields $\varphi'$ and $\varphi$ also vanish (no diagrams for
such functions can be constructed). It is therefore sufficient to consider
only 1-irreducible functions with $N_{\varphi'} \ge 1$ and even sum
$N_{\varphi'} + N_{\varphi}$. Straightforward analysis of the expression
(\ref{IndeX}) then shows that superficial UV divergences can be present
only in the following 1-irreducible functions:
\[ \langle \varphi' \varphi' \rangle \quad (\delta=0) \quad
{\rm with\ the\ counterterm} \quad \varphi' \varphi',  \]
\[ \langle \varphi' \varphi \rangle\, \quad (\delta=2) \quad
{\rm with\ the\ counterterms} \quad \varphi' \partial _{t}\varphi, \quad
\varphi'\partial^{2}_{\parallel} \varphi, \quad
\varphi'\partial^{2}_{\bot} \varphi, \quad
\tau_{0} \varphi' \varphi, \]
\[ \langle \varphi' \varphi^{3} \rangle \quad (\delta=0) \quad
{\rm with\ the\ counterterm} \quad \varphi' \varphi^{3}, \]
\[ \langle \varphi' \varphi v \rangle \quad (\delta=1), \]
for which the counterterm necessarily reduces to the form
$\varphi' (v_{i}\partial_{i}) \varphi = \varphi' v \partial_{\parallel}
\varphi$. All such terms are present in the action (\ref{action}), so that
our model appears multiplicatively renormalizable.

The superficial divergence in the function
$\langle \varphi' \varphi vv \rangle$
with $\delta=0$ and the counterterm $\varphi' \varphi v^{2}$,
allowed by the dimension, is in fact forbidden by the Galilean symmetry.
Furthermore, the latter requires that the counterterms
$\varphi' \partial _{t}\varphi$ and $\varphi' (v_{i}\partial_{i}) \varphi$
enter the renormalized action only
in the form of Lagrangian derivative $\varphi' \nabla _{t}\varphi$.

The arguments based on the Galilean symmetry are usually applied
to the velocity field governed by the Navier--Stokes equation, and
generally become invalid for synthetic Gaussian velocity ensembles.
It turns out, however, that for a Gaussian ensemble with
{\it vanishing} correlation time the Galilean symmetry takes place;
see e.g. \cite{FGV}. This issue, along with the consequences of
the Galilean invariance for the renormalization in our model, are
discussed in \ref{sec:Galileo} in detail.

We conclude that the renormalized action can be written in the form
\begin{eqnarray}
\S_{R}(\varphi',\varphi) = \S_{v} (\bfv) + Z_{1} \sigma (\varphi')^{2} +
\nonumber \\
+  \varphi' \left[
- Z_{2} \sigma \nabla_{t} \varphi + Z_{3} \partial^{2}_{\bot} \varphi +
Z_{4} u \partial^{2}_{\parallel} \varphi - Z_{5} \tau \varphi -
Z_{6} gu^{1/2}\mu^{\eps} \varphi^{3} /6 \right] .
\label{RenAct}
\end{eqnarray}
Here $\sigma$, $\tau$, $u$, $w$ and $g$ are renormalized analogs of the
bare parameters (with the subscripts ``0'') and $\mu$ is the reference mass
scale (additional arbitrary parameter of the renormalized theory).
Since the first term $\S_{v} (\bfv)$ is not renormalized, the amplitude
$D_{0}$ is expressed in renormalized parameters as
\begin{eqnarray}
D_{0} = w_{0} u_{0} /\sigma_{0} = wu\mu^{\xi}/\sigma .
\label{RenD}
\end{eqnarray}

Expression (\ref{RenAct}) is equivalent to the multiplicative
renormalization of the fields $\varphi \to \varphi Z_{\varphi}$,
$\varphi' \to \varphi' Z_{\varphi'}$ and the parameters:
\begin{eqnarray}
\sigma_{0} = \sigma Z_{\sigma}, \quad \tau_{0} = \tau Z_{\tau}, \quad
u_{0} = u Z_{u}, \quad
g_{0} = g \mu^{\eps} Z_{g}, \quad w_{0} = w \mu^{\xi} Z_{w}
\label{Multy}
\end{eqnarray}
(no renormalization of the velocity field is needed: $Z_{v}=1$).
The constants in Eqs. (\ref{RenAct}) and (\ref{Multy}) are related
as follows:
\begin{eqnarray}
Z_{1} = Z_{\sigma} Z^{2}_{\varphi'}, \quad
Z_{2} = Z_{\sigma} Z_{\varphi'} Z_{\varphi}, \quad
Z_{3} = Z_{\varphi'} Z_{\varphi}, \nonumber \\
Z_{4} = Z_{u} Z_{\varphi'} Z_{\varphi}, \quad
Z_{5} = Z_{\tau} Z_{\varphi'} Z_{\varphi}, \quad
Z_{6} = Z_{g}Z_{u}^{1/2} Z_{\varphi'} Z_{\varphi}^{3},
\label{ZZ}
\end{eqnarray}
and from the relation (\ref{RenD}) one obtains:
\begin{eqnarray}
Z_{u}Z_{w}Z_{\sigma}^{-1}=1 .
\label{ZZ1}
\end{eqnarray}

The renormalization constants capture all the divergences at $\eps,\xi\to 0$,
so that the correlation functions of the renormalized model (\ref{RenAct})
have finite limits for $\eps$, $\xi = 0$ when expressed in renormalized
parameters $\sigma$, $\mu$ and so on. In practical calculations, we will use
the minimal subtraction (MS) scheme, in which the renormalization constants
have the forms $Z_{i}=1+\,$ only singularities in $\eps$ and $\xi$, with the
coefficients depending on the two completely dimensionless parameters ---
renormalized coupling constants $g$ and $w$.

\section{RG functions and RG equations} \label{sec:RGE}

Let us recall an elementary derivation of the RG equations; detailed
discussion can be found in monographs \cite{Zinn,Book3}.
The RG equations are written for the renormalized
correlation functions $G_{R} =\langle \Phi\cdots\Phi\rangle_{R}$, which
differ from the original (unrenormalized) ones
$G =\langle \Phi\cdots\Phi\rangle$ only by normalization and choice of
parameters, and therefore can equally be used for analyzing the critical
behaviour. The relation $\S_{R} (\Phi,e,\mu) = \S(\Phi,e_{0})$ between the
functionals (\ref{action}) and (\ref{RenAct}) results in the relations
\begin{equation}
G(e_{0},\dots) = Z_{\varphi}^{N_{\varphi}} Z_{\varphi'}^{N_{\varphi'}}
G_{R}(e,\mu,\dots)
\label{multi}
\end{equation}
between the correlation functions. Here, as usual,
$N_{\varphi}$ and $N_{\varphi'}$ are the numbers of corresponding fields
entering into $\Gamma$ (we recall that in our model $Z_{v}=1$);
$e_{0}=\{\sigma_{0}, \tau_{0}, u_{0}, w_{0}, g_{0} \}$ is the full set of
bare parameters and $e=\{ \sigma, \tau, u, w, g \}$ are their renormalized
counterparts; the dots stand for the other arguments
(times, coordinates, momenta {\it etc}).

We use $\widetilde{\cal D}_{\mu}$ to denote the differential operation
$\mu\partial_{\mu}$ for fixed $e_{0}$ and operate on both sides of the
equation (\ref{multi}) with it. This gives the basic RG differential
equation:
\begin{equation}
\left\{ {\cal D}_{RG} + N_{\varphi}\gamma_{\varphi} +
N_{\varphi'}\gamma_{\varphi'} \right\} \,G^{R}(e,\mu,\dots) = 0,
\label{RG1}
\end{equation}
where ${\cal D}_{RG}$ is the operation $\widetilde{\cal D}_{\mu}$
expressed in the renormalized variables:
\begin{equation}
{\cal D}_{RG}\equiv {\cal D}_{\mu} + \beta_{g}\partial_{g} +
\beta_{w}\partial_{w} - \gamma_{u}{\cal D}_{u} -
\gamma_{\sigma}{\cal D}_{\sigma} - \gamma_{\tau}{\cal D}_{\tau}.
\label{RG2}
\end{equation}
Here we have written ${\cal D}_{x}\equiv x\partial_{x}$ for any variable
$x$, and the anomalous dimensions $\gamma$ are defined as
\begin{equation}
\gamma_{F}\equiv \Dm \ln Z_{F} \quad {\rm for\ any\ quantity} \ F,
\label{RGF1}
\end{equation}
and the $\beta$ functions for the two dimensionless couplings $g$ and $w$ are
\begin{equation}
\beta_{g} \equiv \widetilde {\cal D}_{\mu} g = g\,[-\eps-\gamma_{g}],
\quad
\beta_{w} \equiv \widetilde {\cal D}_{\mu} w = w\,[-\xi-\gamma_{w}],
\label{betagw}
\end{equation}
where the second equalities come from the definitions and the
relations (\ref{Multy}).

Equations (\ref{ZZ}) result in the following relations between the anomalous
dimensions
\begin{eqnarray}
\gamma_{1} = \gamma_{\sigma} + 2 \gamma_{\varphi'} , \quad
\gamma_{2} = \gamma_{\sigma} +  \gamma_{\varphi'}+  \gamma_{\varphi} , \quad
\gamma_{3} = \gamma_{\varphi'}+  \gamma_{\varphi} , \nonumber \\
\gamma_{4} = \gamma_{u} +  \gamma_{3} , \quad \gamma_{5} =
\gamma_{\tau} +  \gamma_{3} , \quad \gamma_{6} =
\gamma_{g}+\gamma_{u}/2+\gamma_{\varphi'}+ 3\gamma_{\varphi} ,
\label{gammas}
\end{eqnarray}
while from (\ref{ZZ1}) one obtains
\begin{eqnarray}
\gamma_{u} + \gamma_{w} -\gamma_{\sigma}=0.
\label{gaas}
\end{eqnarray}
The dimensions $\gamma_{1}$--$\gamma_{6}$ are calculated from the
corresponding renormalization constants using the definition (\ref{RGF1}),
while the RG functions entering equation (\ref{RG2}) are easily found from
the relations (\ref{gammas}) and (\ref{gaas}):
\begin{eqnarray}
2\gamma_{\varphi'}= \gamma_{1}-\gamma_{2}+\gamma_{3}, \quad
2\gamma_{\varphi}= \gamma_{3}-\gamma_{1}+\gamma_{2}, \nonumber \\
\gamma_{u} = \gamma_{4}-\gamma_{3}, \quad
\gamma_{\tau} = \gamma_{5}-\gamma_{3}, \quad
\gamma_{\sigma} = \gamma_{2}-\gamma_{3}, \nonumber \\
\gamma_{w} = \gamma_{2}-\gamma_{4}, \quad
\gamma_{g} = \gamma_{1}-\gamma_{2}-3\gamma_{3}/2 - \gamma_{4}/2+\gamma_{6}.
\label{gammas2}
\end{eqnarray}

The diagrams needed for our second-order calculation of the
critical dimensions are presented in the appendix~B.
One can see that the leading contributions to
different renormalization constants (and hence to the
corresponding anomalous dimensions) are of different order:
$Z_{1,2,3} = 1 +O(g^{2})$, $Z_{5,6}= 1+O(g)$, $Z_{4}= 1+O(w)$.
Practical calculation of the renormalization constants and
anomalous dimensions are discussed in \ref{sec:Graphs},
and here we only present the leading-order results for the
dimensions (\ref{gammas}):
\begin{eqnarray}
\gamma_{1} =\gamma_{2} = b \kappa_{1} \tilde g^{2} +O (\tilde g^{3}),
\quad
\gamma_{3} = \kappa_{1} \tilde g^{2}/6 +O (\tilde g^{3}),
\quad
\gamma_{6} = - 3 \kappa_{2} \tilde g +O (\tilde g^{2}),
\nonumber \\
\gamma_{4} = \tilde w+ \kappa_{1} \tilde g^{2}/6 +O (\tilde g^{3}),
\quad
\gamma_{5} = - \kappa_{1} \tilde g  + O (\tilde g^{2}),
\label{gammaOne}
\end{eqnarray}
where we have denoted $\tilde g = g/(16\pi^2)$, $\tilde w = w/(4\pi^2)$ and
$b=\ln(4/3) \approx 0.287683$; in counting the orders it is assumed that
$w = O(g)$. For generality, we give the results for the $O(N)$-symmetric
model with an $N$-component order parameter in
(\ref{LG}); the additional symmetry factors are $\kappa_{1}=(N+2)/3$ and
$\kappa_{2}=(N+8)/9$. In what follows, we will only give the results for
$N=1$ and denote the new couplings $\tilde g$, $\tilde w$ simply by
$g$, $w$. Then expressions (\ref{gammaOne}) take on the form
\begin{eqnarray}
\gamma_{1} =\gamma_{2} = b g^{2} +O (g^{3}), \quad
\gamma_{3} = g^{2}/6 +O (g^{3}), \quad
\gamma_{6} = - 3 g +O (g^{2}),
\nonumber \\
\gamma_{4} = w+ g^{2} /6 +O (g^{3}), \quad
\gamma_{5} = -  g + O (g^{2}).
\label{gammaTwo}
\end{eqnarray}

\section{Fixed points and scaling regimes} \label{sec:FPS}

It is well known that possible large-scale scaling regimes of a
renormalizable model are associated with IR attractive fixed points of the
corresponding RG equations. In our model, the coordinates $g_{*}$, $w_{*}$
of the fixed points are found from the equations
\begin{equation}
\beta_{g} (g_{*},w_{*}) = 0, \quad \beta_{w} (g_{*},w_{*})=0 ,
\label{points}
\end{equation}
with the $\beta$ functions given in (\ref{betagw}).
The type of a fixed point is determined by the matrix
\begin{equation}
\Omega=\{\Omega_{ij}=\partial\beta_{i}/\partial g_{j}\},
\label{OmegaDef}
\end{equation}
where $\beta_{i}$ denotes the full set of the $\beta$ functions and
$g_{j}= \{g,w\}$ is the full set of couplings. For IR stable fixed points
the matrix $\Omega$ is positive, i.e., the real parts of all its
eigenvalues are positive.

From the definitions (\ref{betagw}), relations (\ref{gammas2}) and explicit
expressions (\ref{gammaTwo}) for the anomalous dimensions we derive the
following leading-order expressions for the $\beta$ functions:
\begin{equation}
\beta_{g} = g\, [-\eps + 3g + w/2], \quad \beta_{w} = w\, [-\xi + w]
\label{betas2}
\end{equation}
with the corrections in the square brackets of order $O(g^{2})$ and higher.
From Eqs. (\ref{points}) and (\ref{betas2}) we can identify four different
fixed points; the matrix $\Omega$ appears triangular for all of them, so
that its eigenvalues are simply given by the diagonal elements
$\Omega_{g} = \partial \beta_{g} / \partial g$ and
$\Omega_{w} = \partial \beta_{w} / \partial w$:

\

1. Gaussian (free) fixed point: $g_{*}=w_{*}=0$;\ \
$\Omega_{g} = -\eps$, $\Omega_{w} = -\xi$.

\

2. $w_{*}=0$ (exact result to all orders), $g_{*}=\eps/3$;
$\Omega_{g} = \eps$, $\Omega_{w} = -\xi$. In this regime, effects of the
velocity field are irrelevant, the isotropy violated by the velocity
ensemble is restored and the leading terms of the IR behaviour coincide
exactly with those of the equilibrium model A. In particular, the basic
critical dimensions do not depend on $\xi$ and coincide to all orders in
$\eps$ with the well-known static exponents $\eta$, $\nu$ for the
Landau--Ginzburg model (\ref{LG}) and the dynamic exponent $z$ for the
model A (see e.g. \cite{Book3,MF}). However, corrections to the
leading-order asymptotic expressions will be anisotropic and different from
those for the model A; in particular, the dependence on $\xi$ will appear
e.g. due to the correction exponent $\Omega_{w}$.

\

3. $g_{*}=0$ (exact result to all orders), $w_{*}=\xi$;
$\Omega_{g} = (-\eps+\xi/2)$, $\Omega_{w} = \xi$. In this regime, the
nonlinearity $\varphi^{3}$ in the stochastic equation (\ref{eq1})
becomes irrelevant, and we arrive at the model of a linear
convection-diffusion equation for a passive scalar field $\varphi$.
For the strongly anisotropic Gaussian velocity ensembles of the type
(\ref{veloc1}), (\ref{veloc2}) such models were investigated in detail
in Refs.~\cite{AM,Glimm,AM2} (mostly for $d=2$, but beyond the scope
of any perturbation theory).

\

4. $g_{*}=(\eps-\xi/2)/3$, $w_{*}=\xi$; $\Omega_{g} = (\eps-\xi/2$),
$\Omega_{w} = \xi$. This fixed point corresponds to a new nontrivial IR
scaling regime, in which the both nonlinearities in the stochastic
equation for $\varphi$ are important; the corresponding critical dimensions
depend essentially on the both RG expansion parameters $\eps$ and $\xi$
and are calculated as double series in these parameters;
see section~\ref{sec:DimeNS}. This
behaviour reveals strong anisotropy and belongs to a new, completely
non-equilibrium, universality class in the sense that the equal-time
correlation functions are not given by a Gibbs measure
$\exp \{- {\cal H} (\varphi) \}$ with a Hamiltonian of the type (\ref{LG}).

\

\begin{figure}
\begin{center}
\includegraphics[width=11cm]{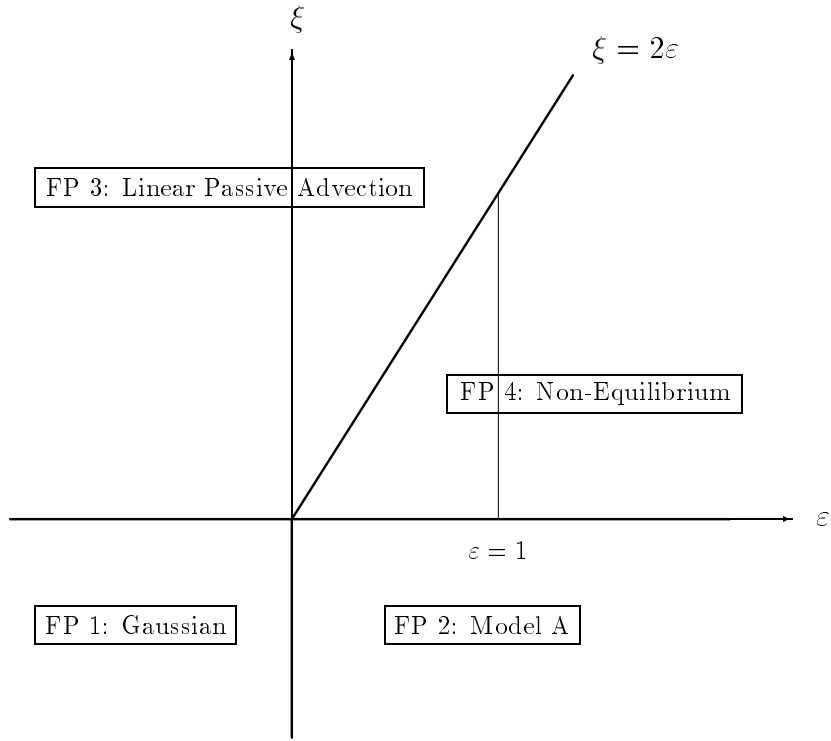}
\caption{\label{fig:pattern} Regions of stability of
the fixed points in the model (\protect\ref{action}).}
\end{center}
\end{figure}

In figure~\ref{fig:pattern} we show the regions of IR stability for all
these fixed points in the $\eps$--$\xi$ plane, that is, the regions in which
the eigenvalues $\Omega_{g,w}$ for a given fixed point are both positive.

In the leading-order approximation (\ref{betas2}), all the boundaries
of the regions of stability are given by straight lines; there are neither
gaps nor overlaps between the different regions. However, experience with
analogous two-parameter models (e.g. double expansion for the stochastic
Navier--Stokes equation near two dimensions \cite{local}) suggests that
such behaviour can rather be an artifact of the leading-order approximation:
the boundaries become curved and overlaps can appear if the higher-order
corrections in the $\beta$ functions are taken into account. In our model
this definitely happens for the boundary between the regions of stability
of the fixed points 2 and 4, as can be argued without practical calculation
of the corrections to the functions (\ref{betas2}); see
section~\ref{sec:DimeNS}.

One can see that the interval of the most realistic values of these
parameters, $\eps=1$ ($d=3$) and $0<\xi<2$ (see the remark above
equation (\ref{effect})), belongs completely to the region of stability
of the most nontrivial fixed point~4. It is also worth noting that,
for all fixed points, the coordinates $g_{*}$, $w_{*}$ are positive in the
regions of their IR stability, in agreement with the physical meaning of
the these parameters: $w$ enters the amplitude in a pair correlation function
and $g>0$ is required for the stability of the static model (\ref{LG}).

\section{Critical scaling and critical dimensions} \label{sec:DimeNS}

Let $F$ be some function of $n$ independent arguments
$\{ x_{1}, \dots, x_{n} \}$ satisfying the following scaling relation
\begin{eqnarray}
F(\lambda^{\alpha_{1}} x_{1}, \dots, \lambda^{\alpha_{n}} x_{n}) =
\lambda^{\alpha_{F}} F(x_{1}, \dots, x_{n})
\label{skal3}
\end{eqnarray}
with a set of constant coefficients (scaling dimensions)
$\{ \alpha_{1}, \dots, \alpha_{n}, \alpha_{F} \}$ and any positive parameter
$\lambda>0$. Differentiating (\ref{skal3}) with respect to $\lambda$ and
setting $\lambda=1$ gives the first-order differential equation
\begin{eqnarray}
\sum_{i=1}^{n} \alpha_{i} {\cal D}_{i} \, F(x_{1}, \dots, x_{n}) =
\alpha_{F} \, F(x_{1}, \dots, x_{n}) , \quad {\cal D}_{i}=x_{i} \partial /
\partial x_{i}
\label{skal1}
\end{eqnarray}
whose general solution has the form
\begin{eqnarray}
F(x_{1}, x_{2}, \dots, x_{n}) = x_{1}^{\alpha_{F}/\alpha_{1}} \,
\widetilde F \left( \frac{x_{2}}{x_{1}^{\alpha_{2}/\alpha_{1}}} \
,\dots,\ \frac{x_{n}} {x_{1}^{\alpha_{n}/\alpha_{1}}} \right),
\label{skal2}
\end{eqnarray}
where $\widetilde F$ is an arbitrary function of $(n-1)$ arguments.
Obviously, the dimensions are determined up to an overall constant factor
(replace $\lambda \to \lambda^{a}$ in (\ref{skal3}) or multiply (\ref{skal1})
by $a$); this arbitrariness can be fixed e.g. by setting $\alpha_{1}=1$. If
$\alpha_{i}=0$ for some $x_{i}$, this variable is not dilated in
(\ref{skal3}) and the corresponding derivative in (\ref{skal1}) is absent.

It is well known that the leading term of the large-scale asymptotic
behaviour of a (renormalized) correlation function satisfies the RG
equation (\ref{RG1}) in which the renormalized couplings are replaced
with their fixed-point values. In our case this gives
\begin{eqnarray}
\left\{ \D_{\mu} - \gamma_{u}^{*} \D_{u} - \gamma_{\sigma}^{*} \D_{\sigma}
-  \gamma_{\tau}^{*} \D_{\tau} + N_{\Phi} \gamma^{*} _{\Phi} \right\}
G_{N_{\Phi}} =0,
\label{RGE}
\end{eqnarray}
where $\gamma_{u}^{*} = \gamma_{u} (g=g_{*}, w=w_{*})$ and so on, and
$G_{N_{\Phi}}$ is the renormalized correlation function with
$N_{\Phi} = \{ N_{\varphi'}, N_{\varphi}, N_{v} \}$ fields. The
summation over all types of fields in the last term of (\ref{RGE}) and
analogous expressions below is implied; cf. equation (\ref{dGamma}).

Canonical scale invariance of the function $G_{N_{\Phi}}$ with respect to
the three independent canonical dimensions (see section~\ref{sec:Reno})
can be expressed by the differential equations of the form
(for definiteness, we consider the momentum--frequency representation)
\begin{eqnarray}
\left\{ \D_{\omega} - \D_{\sigma} - N_{\varphi'} -N_{v} \right\}
G_{N_{\Phi}} =0 ,
\label{CanScal1} \\
\left\{ \D_{\bot} + \D_{\mu} + 2 \D_{\sigma} + 2 \D_{\tau} + 2
\D_{u} - (N_{\varphi'} + N_{\varphi}) (d-3)/2  \right\} G_{N_{\Phi}} =0 ,
\label{CanScal2} \\
\left\{ \D_{\parallel} - 2 \D_{u} - \left( N_{\varphi'} + N_{\varphi}
\right)/2 + N_{v} \right\} G_{N_{\Phi}} =0,
\label{CanScal3}
\end{eqnarray}
where $\D_{\parallel} = k_{\parallel} \partial / \partial k_{\parallel}$,
$\D_{\bot} = k_{\bot} \partial / \partial k_{\bot}$, and the data from
table~\ref{table1} have been used.

Clearly, equation (\ref{RGE}) corresponds to the scaling behaviour
(\ref{skal3}) of the function $G_{N_{\Phi}}$ upon the dilatation of
the parameters $\sigma$, $\mu$, $u$  and $\tau$ and at fixed momentum
and frequency variables; equation (\ref{CanScal1}) deals with dilatation
of the frequencies and other parameters at fixed momenta, and so on. We
are interested in the critical scaling behaviour, that is, behaviour of
the type (\ref{skal3}) in which all the IR relevant parameters
(momenta/coordinates, frequencies/times, deviation of the temperature
from its critical value $\tau \propto (T-T_{c})$) are dilated, while the
IR irrelevant parameters (those which remain finite at the fixed point:
$\sigma$, $\mu$ and $u$) are fixed \cite{Zinn,Book3}. Thus we combine the
equations (\ref{RGE})--(\ref{CanScal3}) so that the derivatives with respect
to the IR irrelevant parameters are eliminated; this gives the desired
equation which describes the critical scaling behaviour:
\begin{eqnarray}
\left\{ \D_{\bot} + \Delta_{\parallel} \D_{\parallel} +
\Delta_{\omega} \D_{\omega} + \Delta_{\tau} \D_{\tau} - N_{\Phi}
\Delta_{\Phi} \right\} G_{N_{\Phi}} =0 .
\label{Krit}
\end{eqnarray}
Here $\Delta_{\bot}=1$ is the normalization condition, while the critical
dimensions of any other IR relevant parameter $F$ is given by the general
expression
\begin{eqnarray}
\Delta_{F} = d_{F}^{\bot}+ \Delta_{\parallel} d_{F}^{\parallel}+
\Delta_{\omega} d_{F}^{\omega} + \gamma^{*}_{F}
\label{KritDim}
\end{eqnarray}
with the canonical dimensions from table~\ref{table1} and
\begin{eqnarray}
\Delta_{\omega}=2+\gamma_{\sigma}^{*} , \quad \Delta_{\parallel} =
\left( 2+\gamma_{u}^{*} \right)/2.
\label{KritDim2}
\end{eqnarray}

Below we will concentrate on the two nontrivial fixed points 2 and~4;
see section~\ref{sec:FPS}.

In the fixed point 2, where $w_{*}=0$ exactly, all the anomalous dimensions
$\gamma^{*}_{F}$ in the RG equation (\ref{RG1}), (\ref{RG2}) coincide with
their counterparts for the equilibrium model A. In particular,
$\gamma_{u}^{*}=\gamma_{4}^{*}-\gamma_{3}^{*}=0$ and therefore
$\Delta_{\parallel}=1$; the $SO(d)$ symmetry violated by the velocity
ensemble is restored. Furthermore, from the fluctuation--dissipation theorem
it follows that $\gamma_{1}^{*}=\gamma_{2}^{*}$; see e.g. \cite{Book3,MF}.
The standard notation for this equilibrium case is
\begin{eqnarray}
\Delta_{\varphi}=d/2-1+\eta/2, \quad \Delta_{\tau}=1/\nu, \quad
\Delta_{\omega}=z,
\label{static}
\end{eqnarray}
while for $\varphi'$ the aforementioned relations for $\gamma^{*}_{F}$ give
$\Delta_{\varphi'}= \Delta_{\varphi}+z=d/2-1+z+\eta/2$. The exponents $\eta$
and $\nu$ can be found directly from the {\it static} model (\ref{LG}); they
are well known from the $4-\eps$, $2+\eps$ and $1/N$-expansions, real-space
RG (all augmented by various summations), high-temperature expansions for the
Ising model (considered the most reliable), Monte-Carlo simulations. The
values recommended by \cite{Zinn,Book3} are $\eta=0.0375 \pm 0.0025$ and
$\nu= 0.6310\pm0.0015$ (Borel summation of 5-order results). For $z$ only
two terms of the $4-\eps$ expansion are known:
$z= 2+0.726 (1-0.1885 \eps)\,\eta$ \cite{AV}; there are also four-loop
results in the real-space RG \cite{Prudnikov} and leading-order results
in $2+\eps$ and $1/N$-expansions; see the discussion in \cite{Book3,HH,MF}
and references therein.

Let us turn to the fixed point~4. We are going to find the critical
dimensions entering the equations (\ref{Krit}) and (\ref{KritDim2})
to the second order of the generalized $\eps$ expansion,
that is, the double expansion in $\eps$ and $\xi$ with the
convention that $\xi=O(\eps)$. From the relations (\ref{gammas2})
and (\ref{gammaTwo}) it follows that
$\gamma_{\sigma,\varphi,\varphi'}=O(g^{2})$ and $\gamma_{u}=O(w)$,
so that in order to find $\gamma_{\sigma,\varphi,\varphi'}^{*}$
with the accuracy of $O(\eps^{2})$ it is sufficient to calculate the
coordinate $g_{*}$ only to the leading order $O(\eps)$. At first sight,
the second-order calculation of the coordinates $g_{*}$, $w_{*}$ is
needed to find the $O(\eps^{2})$ contribution in the dimension
$\gamma_{u}^{*}$. However, this calculation can be avoided with the aid
of the exact identity
\begin{equation}
\gamma_{u}^{*} = \xi + \gamma_{\sigma}^{*},
\label{wne0}
\end{equation}
which follows from the relations $\beta_{w}  = w\,[-\xi-\gamma_{w}]$
in (\ref{betagw}) and $\gamma_{u} + \gamma_{w} = \gamma_{\sigma}$ in
(\ref{gaas}) for any fixed point at which $\beta_{w}=0$ and $w_{*} \ne 0$.
Then from (\ref{KritDim2}) it follows that
\begin{eqnarray}
2 \Delta_{\parallel}= \Delta_{\omega} + \xi
\label{Krit3}
\end{eqnarray}
exactly, and from the explicit expressions (\ref{gammaTwo}) one obtains
($\bar\eps\equiv \eps-\xi/2$):
\begin{eqnarray}
\Delta_{\varphi} = 1 - \bar\eps/2 + (6b+1)\bar\eps^{2}/486,
\nonumber \\
\Delta_{\varphi'} = 3 - \bar\eps/2 + (10b-1)\bar\eps^{2}/72,
\nonumber \\
\Delta_{\parallel} =  1 + \xi/2 + (6b-1)\bar\eps^{2}/108,
\nonumber \\
\Delta_{\omega} =  2 + (6b-1)\bar\eps^{2}/54,
\nonumber \\
\Delta_{\tau} = 2 - \bar\eps/3
\label{Anis}
\end{eqnarray}
with corrections of order $O(\eps^{2})$ for $\Delta_{\tau}$ and $O(\eps^{3})$
for the other dimensions; we recall that $b=\ln(4/3) \approx 0.287683$.

For the velocity field, relations (\ref{KritDim}), (\ref{KritDim2}) and
(\ref{wne0}) can be combined to give the exact expression
\begin{eqnarray}
\Delta_{v} = 1+\gamma_{\sigma}^{*}- \gamma_{u}^{*}/2 =
(\Delta_{\omega}-\xi)/2 ,
\label{DeltaV}
\end{eqnarray}
in agreement with the explicit factorized form of the
velocity correlation function (\ref{veloc1}), (\ref{veloc2}).

It remains to note that the critical dimensions (\ref{Anis})
coincide up to the order $O(\eps)$ with their counterparts in
(\ref{static}) at the ray $\eps>0$, $\xi=0$, the boundary between
the regions of stability of the corresponding fixed points 2 and 4
(determined in section \ref{sec:FPS} in the first-order
approximation for the $\beta$ functions), but differ in order
$O(\eps^{2})$. This is a clear indication that a straight boundary
without gaps and overlaps is an artifact of the first-order
approximation: the boundaries become curved and overlaps or gaps
appear when the higher-order corrections to the $\beta$ functions
are taken into account, as it happens in the analogous double
expansion for the stochastic Navier--Stokes equation near two
dimensions \cite{local}.

\section{Discussion and conclusion} \label{sec:Conc}

We have studied effects of turbulent mixing and stirring on the critical
behaviour of a fluid system (binary mixture, nematic liquid crystal) with
a purely relaxational dynamics of a non-conserved order parameter, known
as model A \cite{Book3}--\cite{MF}. The velocity was modelled by Gaussian
statistics with vanishing correlation time and strongly anisotropic
correlation function $\propto\delta(t-t') /k_{\bot}^{d-1+\xi}$; see
equations (\ref{veloc1}), (\ref{veloc2}). Such ensembles were employed
earlier in \cite{AM,Glimm,AM2} in the analysis of the two-dimensional
passive turbulent advection (linear equation for the scalar field).

The model, originally described by a stochastic differential equation
(\ref{eq1})--(\ref{nabla}), can be reformulated as a multiplicatively
renormalizable field theory (\ref{action}), which allows one to apply the
field theoretic RG to study its critical behaviour. The model reveals four
different IR scaling regimes, related with the four different fixed
points of the RG equations; their regions of stability in the $\eps$--$\xi$
plane are identified in the leading order. These regimes correspond to:
(1) Gaussian (free) model, (2) equilibrium critical dynamics (standard
universality class of the model A, interaction with the velocity field is
irrelevant), (3) linear passive scalar advection (the $\varphi^{4}$ term
in the Landau--Ginzburg Hamiltonian is irrelevant) and (4) the most
nontrivial strongly anisotropic scaling regime in which the both
interactions are important; it corresponds to a new non-equilibrium
universality class.

It was shown that the equilibrium critical regime (model A) becomes
unstable for the realistic range of parameters $d<3$ and $0<\xi<2$,
which includes Kolmogorov spectrum ($\xi=4/3$) and Batchelor limit
($\xi=2$). It is replaced with the new non-equilibrium regime;
the corresponding critical exponents
are calculated to second order of the corresponding RG expansion, which
in this case takes on the form of the double expansion in $\eps$ and $\xi$;
explicit expressions are given in (\ref{Anis}).

Let us discuss the consequences of the general scaling relations, derived
in section \ref{sec:DimeNS}, for the most interesting special case of the
pair correlation function. They result in the scaling expression
\begin{eqnarray}
\big\langle \varphi (\x+\bfr, t+t')\, \varphi (\x, t') \big\rangle =
r_{\bot}^{-2\Delta_{\varphi}} \,
{\cal F} \left( \tau_{0}\,{r^{\Delta_{\tau}}}, \,
t/r_{\bot}^{\Delta_{\omega}}, \,
r_{\parallel}/ r_{\bot}^{\Delta_{\parallel}} \right),
\label{scaling1}
\end{eqnarray}
where  $r_{\bot}=|\bfr_{\bot}|$,  $r_{\parallel}=|\bfr_{\parallel}|$ and
${\cal F}$ is some scaling function. It is usually assumed that ${\cal F}$
has finite limits for $\tau_{0}\propto (T-T_{c})=0$ (that is, exactly at the
critical point) and/or for $t=0$ (equal-time correlation function).
Then from (\ref{scaling1}) one obtains
\begin{eqnarray}
\big\langle \varphi(\x+\bfr,t)\, \varphi(\x,t)\big\rangle
= r_{\bot}^{-2\Delta_{\varphi}} \, \widetilde{\cal F}
\left( r_{\parallel}/ r_{\bot}^{\Delta_{\parallel}} \right)
\label{scaling2}
\end{eqnarray}
with another nontrivial function $\widetilde{\cal F}(x)={\cal F}(0,0,x)$.
The two last arguments in the scaling representation (\ref{scaling1}) can
also be chosen in the form $r_{\bot}/ L_{\bot} (t)$ and $r_{\parallel}
/L_{\parallel} (t)$ with two different characteristic length scales
\begin{eqnarray}
L_{\bot} (t) \sim t^{\alpha_{\bot}}, \quad
L_{\parallel} (t) \sim t^{\alpha_{\parallel}}, \qquad
\alpha_{\bot} = 1/\Delta_{\omega}, \quad
\alpha_{\parallel} = \Delta_{\parallel}/\Delta_{\omega},
\label{scales}
\end{eqnarray}
with the exact relation $2\alpha_{\parallel} = 1+\xi \alpha_{\bot}$
following from equation (\ref{Krit3}). For the most realistic values
$\eps=1$ ($d=3$) and $\xi=4/3$ (Kolmogorov spectrum of the velocity)
explicit results (\ref{Anis}) give
\begin{eqnarray}
\Delta_{\omega} \approx 2.0015, \quad \alpha_{\bot} \approx 0.4996 \quad
{\rm and} \quad \alpha_{\parallel} \approx 0.833,
\label{scales2}
\end{eqnarray}
while for $\eps=1$ and $\xi=2$ (Batchelor limit, smooth velocity field)
in the same approximation one obtains
\begin{eqnarray}
\Delta_{\omega} = 2, \quad \alpha_{\bot} = 0.5 \quad {\rm and} \quad
\alpha_{\parallel} = 1,
\label{scales3}
\end{eqnarray}
with possible
corrections from the $O(\eps^{3})$ terms in (\ref{Anis}). It is worth
noting that the $O(\eps^{2})$ contributions to these results are almost
negligible, so that $\Delta_{\omega}$ appears almost indistinguishable
from its canonical value $\Delta_{\omega}=2$. On the contrary, the analog
of Fisher's exponent $\eta=\xi/2+(6b+1)\bar\eps^{2}/243$, determined from
(\ref{Anis}) using the ``equilibrium'' relation (\ref{static}), markedly
deviates from its canonical (vanishing) value due to the $O(\xi)$ term:
$ \eta\approx 2/3$ for $\xi=4/3$ and $ \eta\approx 1$ for $\xi=2$. This
is reminiscent of the observation made in Refs. \cite{Beysens,Akira}
(however, for a conserved order parameter and a non-random velocity)
that the critical fluctuations are suppressed by the flow and the behaviour
of the system becomes close to the mean-field limit in a strong shear; see
also discussion in \cite{Chan}.

Existence of two different length scales (\ref{scales}) with power-law
dependence on the time was established in a number of studies within
numerical simulations \cite{Shear2,Shear6}, approximate analytical solutions
\cite{Shear3} and exactly soluble simplified models \cite{Shear1}. As a rule,
those authors dealt with binary mixtures in the coexistence (two-phase)
region (finite and negative $\tau_{0} \propto (T-T_{c}) $) in the presence
of a uniform laminar shear flow, while our results refer to a system near
its critical point and in a chaotic velocity ensemble. [For finite
$\tau_{0}<0$, phase separation occurs at length scales comparable to the
typical size of turbulent eddies, the situation which is much more difficult
to achieve in practice in the vicinity of the critical point
($\tau_{0} \simeq 0$), at least for binary mixtures; see also the
discussion in Ref. \cite{Ronis}.] Thus {\it a priori}
one should not have expected a good quantitative agreement for the exponents
in (\ref{scales}). Surprisingly enough, our answers (\ref{scales2}) and
(\ref{scales3}) for the exponents appear not inconsistent with the results
$\alpha_{\bot} =0.5$ and $\alpha_{\parallel} = 3/2$, derived earlier in Refs.
\cite{Shear3} for a non-conserved order parameter in a uniform non-random
shear within the so-called Ohta--Jasnow--Kawasaki approximation \cite{OJK}.
Although the values of $\alpha_{\parallel}$ are rather different in
\cite{Shear3} and (\ref{scales2}), (\ref{scales3}), they are always
markedly larger than $\alpha_{\bot}$. The same inequality
$\alpha_{\parallel} > \alpha_{\bot}$ for the exponents was also
established in two dimensions \cite{Shear3} and for exactly soluble models
\cite{Shear1}.

Let us briefly discuss the general case (\ref{vello}) with ${\bf u} \ne 0$.
Nonvanishing mean velocity ${\bf u}$ gives rise to the additional term
$\sigma_{0}\varphi' (u_{i}\partial_{i}) \varphi =
\sigma_{0}\varphi' (u\partial_{\parallel}) \varphi$ with $u=|\bfu|$
in the action (\ref{action}), which simply results in the replacement
$\omega \to \omega - \bfu\cdot\k = \omega - uk_{\parallel}$ in the
propagators (\ref{lines3}), (\ref{lines2}) and
$r_{\parallel} \to r_{\parallel} + ut$ in the final scaling expressions
like (\ref{scaling1}). This fact can be compared with the observation made
in Refs. \cite{Shear6,Shear3} that, for ${\bf u} \ne 0$, the
proper scaling variables are not simply related to parallel and
perpendicular directions. The dependence on ${\bf u}$ disappears at $t=0$,
that is, in the equal-time correlation function (\ref{scaling2}).

Another interesting quantity is the ``crossover exponent'' $\chi$ in the
relation $\delta T_{c} \propto Re^{\chi}$ between the Reynolds number $Re$
and the shift $\delta T_{c}$ of the critical temperature due to the mixing,
with experimental estimates $\chi \sim 1.4$--$2.1$ \cite{Pine}. In the RG
framework, this exponent can be identified \cite{Satten,Ronis} as
$\chi = \nu |\Omega_{\rm min}|$, where $\nu \simeq 0.63$ is the classical
critical exponent (\ref{static}) for the Landau--Ginzburg model (\ref{LG})
and $\Omega_{\rm min}$ is the minimal (maximal by the modulus) negative
eigenvalue of the $\Omega$ matrix (\ref{OmegaDef}) at the equilibrium
scaling regime (model A or, in our notation, fixed point 2). In our case
$\Omega_{\rm min}= \Omega_{w} = -\xi$ for the point 2; see
section~\ref{sec:FPS}. This gives
$\chi \simeq 1.2$ for the Kolmogorov spectrum ($\xi=4/3$) and $\chi \simeq
1.26$ for the Batchelor limit, which is better than the estimate $\chi \simeq
0.8$ obtained in \cite{Akira} for strongly anisotropic non-random shear but
worse than the RG result $\chi \simeq 1.74$ obtained in \cite{Satten,Ronis}
for a random isotropic velocity ensemble with the velocity spectrum
$\propto 1/k^{2}$. Of course, the disagreement can be explained
by the non-conservation of the order parameter in our model.
[In this connection it should be mentioned that the discussion of section
VI in \cite{Ronis} for the general exponent in the velocity correlation
function, denoted as $1/k^{2+a\eps}$ in equation (6.1) of \cite{Ronis},
contains an error: the second $\beta$ function in (6.8) must be
$\beta_{\lambda}=-\widetilde\lambda_{R}\left\{(1+a)\eps - \dots\right\}$.
Thus the conclusions made in the following discussion about the independence
of the critical exponents on $a$ (in our notation, on the relation between
the two RG expansion parameters $\eps$ and $\xi$) can be erroneous and
must be revisited.]

It remains to note that for $d<3$ and not too small $\xi$, the fixed point
4 becomes unstable while the point 3 becomes IR attractive (see
figure~\ref{fig:pattern}), the $\varphi^{4}$ interaction in (\ref{LG})
becomes irrelevant, and the IR behaviour of the model coincides with that
of the linear passive scalar advected by the anisotropic Gaussian velocity
ensemble (\ref{veloc1}), (\ref{veloc2}). For $d=2$, this regime was
investigated in detail in Refs. \cite{AM,Glimm,AM2}.

We may conclude that our simplified model of a non-conserved order parameter
and Gaussian velocity ensemble captures important characteristics of a real
second-order phase transition in a stirred fluid system: persistence of a
critical scaling regime; emergence of a new non-equilibrium universality
class with a new set of critical exponents, rather different from the
classical ones; existence (for a strongly anisotropic velocity ensemble) of
two different length scales (with a power-law time dependence), and so on.
Further investigation should take into account conservation of the order
parameter and its interaction with other thermodynamical degrees of freedom
(mode-mode coupling), compressibility, non-Gaussian character and finite
correlation time of the velocity field, and so on. This work is now in
progress.

\section*{Acknowledgments}
The authors thank L Ts Adzhemyan, Michal Hnatich and Juha Honkonen for
discussions. The work was supported in part by the Russian Foundation
for Fundamental Research (grant No~05-02-17\,524), the Russian National
Program (grant No~2.1.1.1112) and the program ``Russian Scientific Schools''
(grant No~5538.2006.2). The diagrams were plotted with the aid of JaxoDraw
\cite{Jax}.

\appendix
\section{Consequences of the Galilean symmetry} \label{sec:Galileo}

In this Appendix we will explore consequences of the Galilean symmetry for
the renormalization of the model (\ref{action}). The models with synthetic
Gaussian velocity ensembles are, as a rule, not invariant with respect to
the Galilean transformations. Nevertheless, if the velocity is not correlated
in time, nontrivial parts of the 1-irreducible correlation functions appear
invariant (more precisely, see below), and the Galilean symmetry can be used
to restrict the form of the counterterms. In this sense, the symmetry of the
counterterms is higher than the symmetry of the action functional.

For the most of the following discussion, precise form of the nonlinearity
in (\ref{eq1}) is unessential; it is only important that it is consistent
with Galilean symmetry. The velocity field will be taken divergence-free,
Gaussian, with zero mean and the correlator
\begin{eqnarray}
\langle v_{i}(t,\bfx) v_{j}(t',\bfx') = \delta(t-t') D_{ij}(\bfx-\bfx').
\label{RF}
\end{eqnarray}
Our model (\ref{veloc1}), (\ref{veloc2}) corresponds to a special choice
of the function $D_{ij}$, but in what follows its precise form is also
unessential.

Field theoretic formulation (\ref{action}) means that the generating
functionals of total $[G(A)]$ and connected $[W(A)]$ correlation functions
of the original stochastic problem can be represented by the functional
integral of the form
\begin{eqnarray}
G(A) = \exp W(A) = \int {\cal D} \Phi \, \exp \{ \S(\Phi) + A\Phi \}.
\label{GW}
\end{eqnarray}
Here and below, we denote by $\Phi = \{ \varphi', \varphi, {\bf v} \}$
the full set of fields and by $A = \{ A_{\varphi'},
A_{\varphi} , {\bf A}_{v} \}$ the full set of sources; in the
expressions like
\[ A\Phi = \sum_{\Phi} \int dx   A(x) \Phi(x) \]
summation over all types of the fields, integration over their
arguments $x=\{t,\bfx\}$ and summation over their vector indices
are always understood. All the normalization factors are included
into the functional differential ${\cal D} \Phi = {\cal
D}\varphi'\, {\cal D}\varphi\, {\cal D}\bfv$; the normalization
$G(0)=1$ is implied.

The Galilean transformation is defined as
\[ v_{i}(t,\bfx) \to \widetilde v_{i}(t,\bfx)
= v_{i}(t,\bfx+\bfu t) - u_{i} \]
for the velocity and
\begin{eqnarray}
\Phi (t,\bfx) \to \widetilde \Phi (t,\bfx) = \Phi (t,\bfx+\bfu t)
\label{GT}
\end{eqnarray}
for the other fields; here $\bfu$, the parameter of the transformation,
is an arbitrary constant vector. For the strongly anisotropic ensemble
(\ref{veloc1}), (\ref{veloc2}), the vector $\bfu$ must be parallel to
$\bfv \sim \n$; then the shift of the arguments in (\ref{GT}) reduces to
$x_{\parallel}\to x_{\parallel}+u$.

The part of the action (\ref{action}) which corresponds to the stochastic
problem (\ref{eq1}) at fixed ${\bf v}$ is clearly invariant: one has to
substitute $\Phi \to \widetilde \Phi$ and make the change of variables
$\bfx+\bfu t \to \bfx$; the additional terms
$\varphi' (u_{i}\partial_{i}) \varphi$ coming from the contribution with
$\partial_{t}$ and from the nonlinearity cancel each other in the covariant
combination $\varphi' \nabla_{t} \varphi$. If the velocity were governed by
the stochastic Navier--Stokes equation with a time-decorrelated random force,
the total action would also be invariant (see e.g. \cite{Book3}), but for
our synthetic ensemble (\ref{RF}) variation of the action
$\S_{v} (\bfv)$ in (\ref{Sv}) is nontrivial:
\begin{eqnarray}
\S_{v} (\widetilde\bfv) = \S_{v} (\bfv) + u D^{-1} v + O(u^{2}).
\label{Var}
\end{eqnarray}
In the detailed notation,
\[ u D^{-1} v = \int dt \int d\bfx  \int d\bfx' \
 u_{i} D^{-1}_{ij} (\bfx-\bfx') v_{j}(t,\bfx'), \]
where $D^{-1}$ the inverse linear operation for $D$ in (\ref{RF}) on
the transverse subspace.

We stress that for the validity of (\ref{Var}) it is crucial that the
correlator (\ref{RF}) involves the $\delta$ function in time. Indeed,
substitution $\bfv\to\widetilde\bfv$ in $\S_{v}$ produces the term
\begin{eqnarray}
\int dt \int d\bfx  \int d\bfx' \ v_{i}(t,\bfx+\bfu t) D^{-1}_{ij}
(\bfx-\bfx') v_{j}(t,\bfx'+\bfu t),
\label{Shift}
\end{eqnarray}
which gives $\S_{v}(\bfv)$ after the change of variables
$\bfx+\bfu t \to \bfx$, $\bfx'+\bfu t \to \bfx'$
due to the fact that the both fields in (\ref{Shift}) have the same time
argument and therefore the argument of $D^{-1}$ remains unchanged. For a
finite correlation time, expression (\ref{Shift}) would involve
the double time integral, the argument of $D^{-1}$ would be shifted by
$\bfu(t-t')$ and the original action $\S_{v}$ in the right-hand side of
(\ref{Var}) would not be formed.

Let us make the substitution $\Phi \to \widetilde \Phi$ in the functional
integral (\ref{GW}). This is just a change of integration variables, its
Jacobian equals unity, so the integral
\begin{eqnarray}
G(A)= \int {\cal D} \Phi \, \exp \{ \S(\widetilde\Phi) + A\widetilde\Phi \}
\label{GWG}
\end{eqnarray}
is in fact independent of the parameter $\bfu$ from (\ref{GT}). In particular,
this means that its first variation with respect to $\bfu$ vanishes. Let us
denote by $\delta_{u} F(\Phi)$ the linear-in-$\bfu$ term in the Galilean
transformed quantity $F(\widetilde\Phi)$. Then we have
$\delta_{u} \bfv = (u\partial)\bfv-\bfu$ for the velocity,
$\delta_{u} \Phi = (u\partial)\Phi$ for the other fields and
$\delta_{u} \S(\Phi)=\delta_{u} \S_{v} (\bfv)  = u D^{-1} v $ for the
action functional. Substituting these expressions into the first variation
of (\ref{GWG}) gives the identity
\begin{eqnarray}
\int {\cal D} \Phi \, \big\{ A(u\partial) \Phi -uA_{v} +u D^{-1} v \big\}
\exp \{ \S(\Phi) + A\Phi \} = 0
\label{Iden}
\end{eqnarray}
with implied summations over all types of fields, integrations and so
on. The fields can be taken outside the integral in (\ref{Iden}) as
variational derivatives with respect to the corresponding sources,
$\Phi \to \delta / \delta A$, which gives for the functional $W(A)$ from
(\ref{GW}) the following differential equation in variational derivatives:
\begin{eqnarray}
A(u\partial) \frac{\delta W(A)} {\delta A} -uA_{v} +
u D^{-1} \frac{\delta W(A)} {\delta A_{v}} =0.
\label{VDE}
\end{eqnarray}

It is well known that the generating functional $\Gamma(\Phi)$ of the
1-irreducible correlation functions (sometimes referred to as ``effective
action'') is obtained from $W(A)$ as the functional Legendre transform
with respect to the sources $A$ (see e.g. \cite{Zinn,Book3}):
\begin{eqnarray}
\Gamma(\Phi)= W(A)-A\Phi, \quad \frac{\delta W(A)} {\delta A}=\Phi, \quad
\frac{\delta \Gamma(\Phi)} {\delta \Phi}= -A.
\label{Le}
\end{eqnarray}
Here the sources are (implicitly) expressed as functions of the fields
for a given $W(A)$ using the second relation, while the third relation
explicitly determines $A$ in terms of $\Phi$ for a given $\Gamma(\Phi)$.
Substituting (\ref{Le}) into (\ref{VDE}) gives the following equation for
$\Gamma(\Phi)$:
\begin{eqnarray}
\frac{\delta \Gamma(\Phi)} {\delta \Phi} (u\partial) \Phi -u \frac
{\delta \Gamma(\Phi)} {\delta v}   = u D^{-1} v.
\label{VDG}
\end{eqnarray}
The left-hand side obviously represents the first variation
$\delta_{u} \Gamma(\Phi)$ of the functional (\ref{Le}) with respect to the
Galilean transformation (\ref{GT}) of its functional arguments, while the
right-hand side is nothing but the variation $\delta_{u} \S(\Phi)$ of the
action (\ref{action}). It is well known that the functional (\ref{Le}) can
be represented as the sum $\Gamma(\Phi)= \S(\Phi)+\bar \Gamma(\Phi)$ of the
action $\S(\Phi)$, which contains all the tree (``loopless'') graphs and the
terms not represented by graphs, and the nontrivial part $\bar \Gamma(\Phi)$
which contains all the graphs with loops (and hence those with all possible
UV divergences); see e.g. \cite{Zinn,Book3}. From (\ref{VDG}) we conclude
that the total non-invariance of $\Gamma(\Phi)$ is brought about by the
action term, while the second contribution appears invariant:
$\delta_{u} \bar \Gamma(\Phi)=0$.

The last relation holds for arbitrary values of the model parameters,
including $d$, $\xi$ and the coupling constants (\ref{g0}). Therefore
it remains valid in the perturbation theory and is preserved by the
renormalization procedure. We thus may conclude that the contribution of
the counterterms (determined by the nontrivial term $\bar \Gamma(\Phi)$)
must also be Galilean invariant, in spite of the fact that the total
functional $\Gamma(\Phi)$ is not. This justifies the statements made in
the analysis of the renormalization of our model in section~\ref{sec:Reno}:
the counterterm $\varphi' \varphi v^{2}$, allowed by the dimension, is
not invariant and therefore it is forbidden; the counterterms
$\varphi' \partial_{t}\varphi$ and $\varphi' (v_{i}\partial_{i}) \varphi$
can appear only in the form of the Galilean covariant combination
$\varphi' \nabla _{t}\varphi$.

\section{Calculation of the Feynman diagrams} \label{sec:Graphs}

\begin{figure}
\begin{center}
\includegraphics[width=11cm]{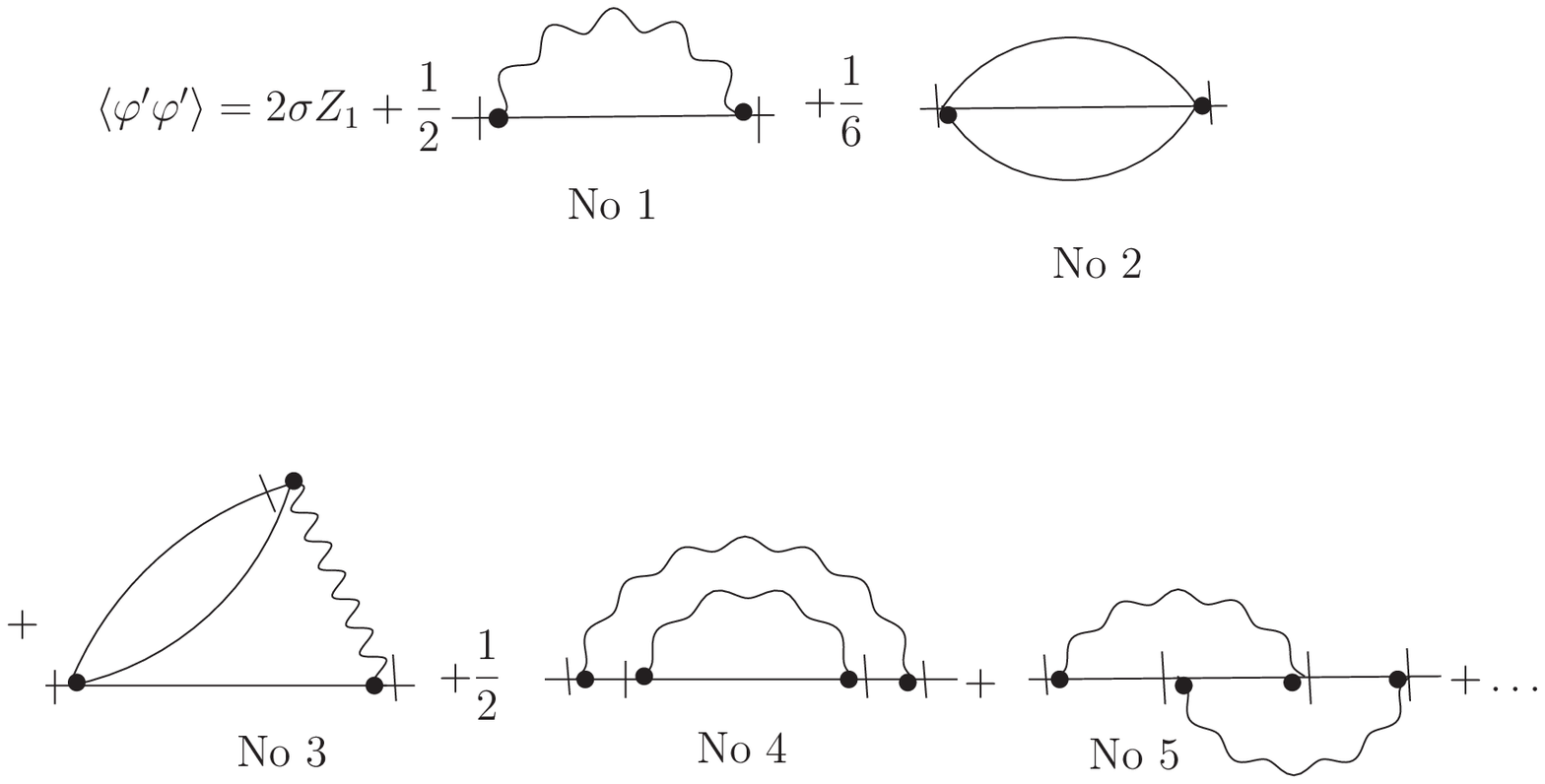}
\caption{\label{Fig2} 1-irreducible function
$\langle\varphi'\varphi'\rangle$: Relevant one- and two-loop
diagrams.}
\end{center}
\end{figure}

\begin{figure}
\begin{center}
\includegraphics[width=11cm]{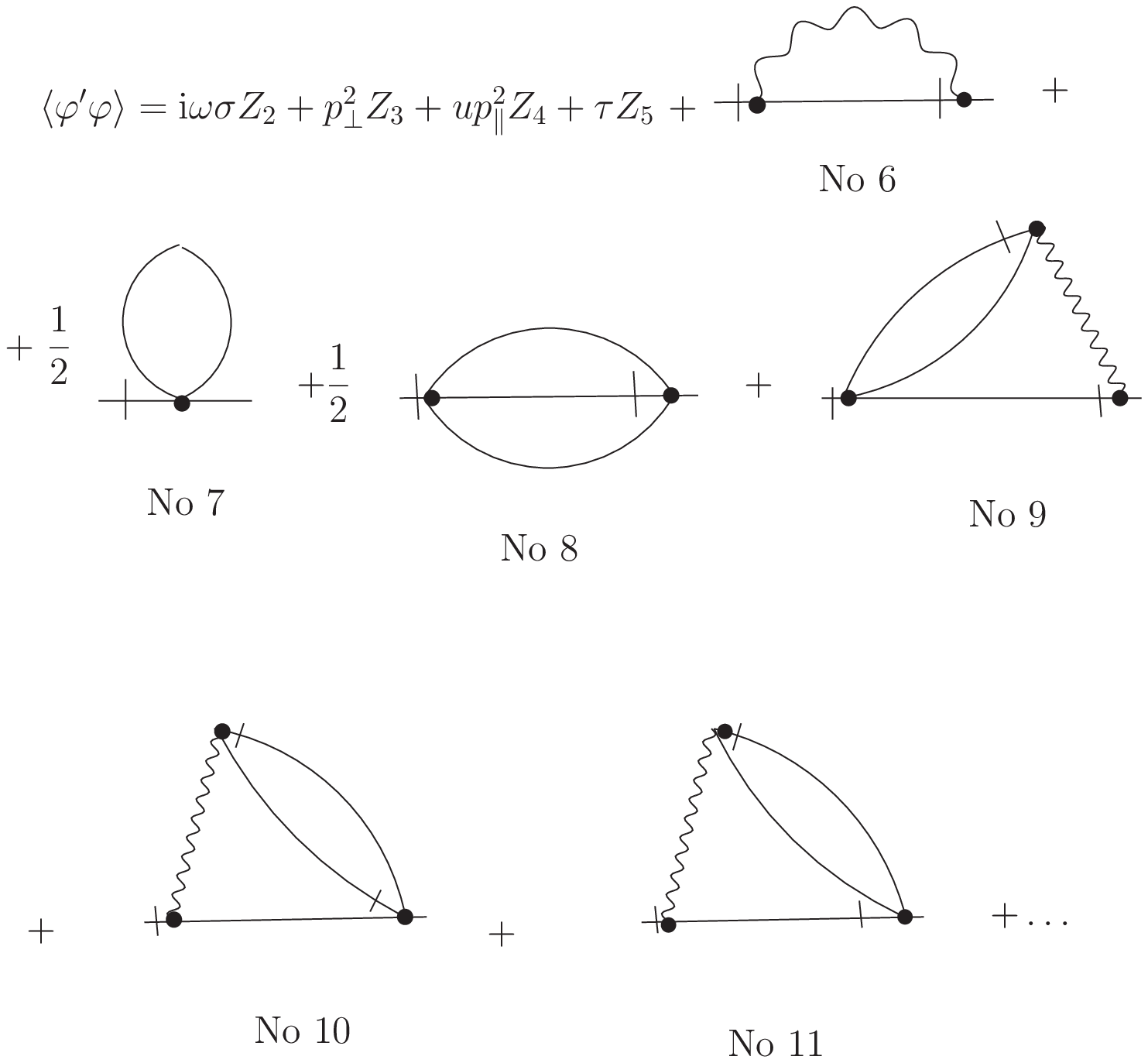}
\caption{\label{Fig3} 1-irreducible function
$\langle\varphi'\varphi\rangle$: Relevant one- and two-loop diagrams.}
\end{center}
\end{figure}

\begin{figure}
\begin{center}
\includegraphics[width=11cm]{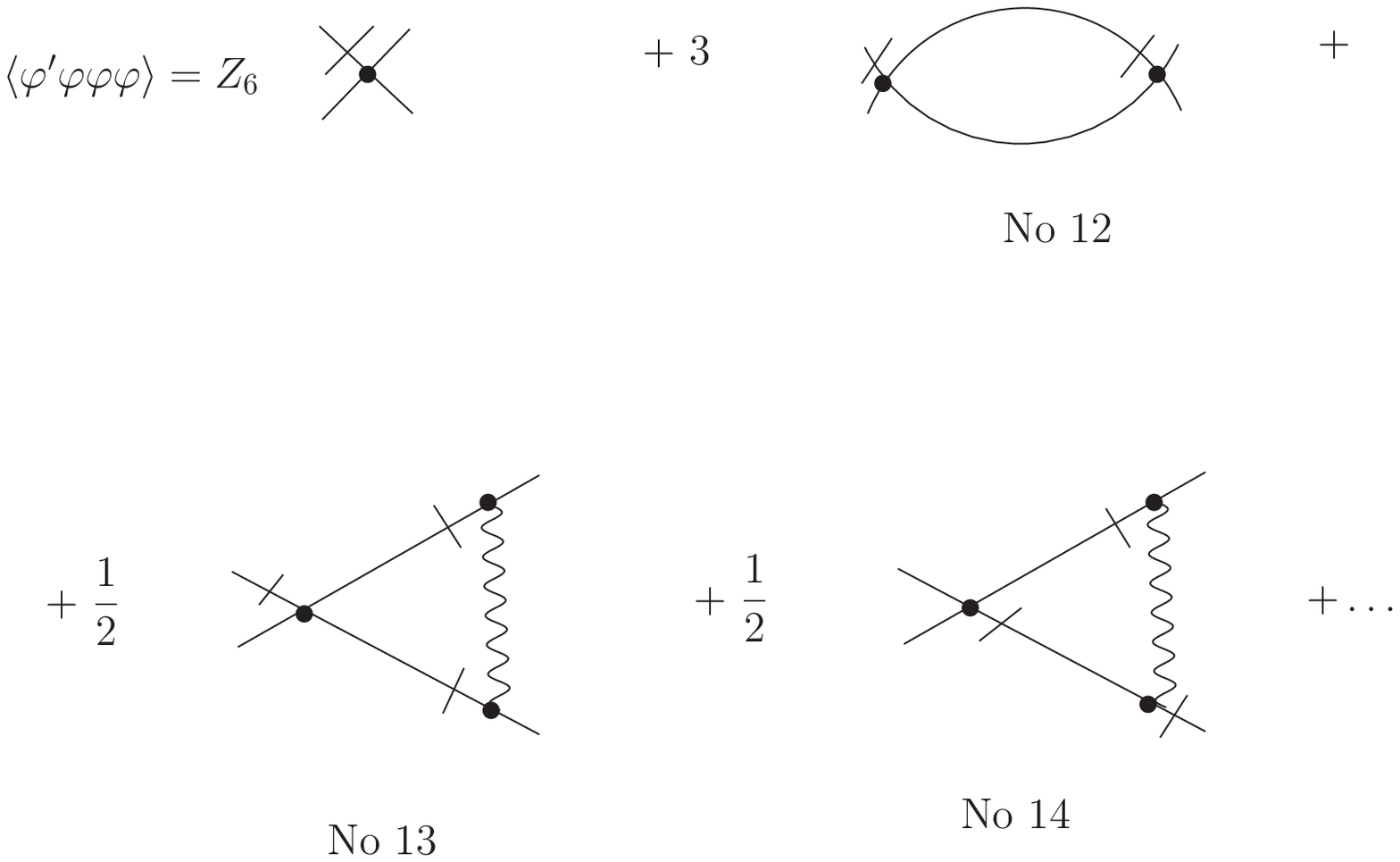}
\caption{\label{Fig4} 1-irreducible function
$\langle\varphi'\varphi\varphi\varphi\rangle$: One-loop
approximation.}
\end{center}
\end{figure}

In this appendix we will briefly discuss the main points
concerning the calculation of the renormalization constants and
the corresponding Feynman diagrams. In order to find the anomalous
dimensions (\ref{gammas}) in the approximation (\ref{gammaOne}),
(\ref{gammaTwo}), one has to calculate the 1-irreducible
correlation functions $\langle \varphi' \varphi' \rangle$,
$\langle \varphi' \varphi \rangle$ to the two-loop order and
$\langle \varphi' \varphi\varphi\varphi \rangle$ to the one-loop order of
the renormalized perturbation theory. The corresponding
diagrammatic expressions are given in figures \ref{Fig2}--\ref{Fig4}
(we do not show some of the diagrams, which are topologically possible
but vanish because of special reasons; see below).

The wavy lines denote the pair correlator of the velocity (\ref{veloc2}).
The bare propagators (\ref{lines3}), (\ref{lines2}) are denoted by solid
lines: the line without a slash denotes the propagator
$\langle \varphi \varphi \rangle_{0}$ while the line with one slash
denotes $\langle \varphi \varphi' \rangle_{0}$; the slashed ends correspond
to the response field $\varphi'$. The propagator
$\langle \varphi' \varphi' \rangle_{0}$
vanishes in any dynamic model of the type (\ref{action}), so the
solid line with two slashed ends does not occur in the diagrams.
There are two types of vertices (shown by thick dots):
the quartic one corresponds to
the interaction $\varphi' \varphi^{3}$ and the triple one
corresponds to $\varphi' (v\partial_{\parallel}) \varphi$. The
symmetry coefficients are shown for the one-component scalar field
($N=1$).

All the diagrammatic elements should be expressed in renormalized variables
using the relations (\ref{RenAct})--(\ref{ZZ1}). However, in our
approximation the $Z$'s should be retained (with the appropriate accuracy
in $g$ and $w$) only in the bare terms of the functions
$\langle \Phi\dots \Phi \rangle$, while in the diagrams they should be
replaced with unities. In other words, the passage to renormalized
variables in the diagrams is provided by the simple substitutions
$\sigma_{0} \to \sigma$, $u_{0} \to u$, $\tau_{0} \to \tau$,
$g_{0} \to g\mu^{\eps}$ and $w_{0} \to w\mu^{\xi}$.

In the practical calculation we use the MS scheme, where the
renormalization constants are independent of the specific choice of the IR
regularization. It is then possible to calculate the constants
directly in the critical (``massless'') model, that is, at $\tau=0$
in the renormalized analogs of the lines (\ref{lines3}), (\ref{lines2}).
Then, in the calculation of the constants $Z_{i}$ with $i\ne5$,
the diagrams involving self-contracted propagators
$\langle \varphi \varphi \rangle_{0}$ can be treated as zero and, for this
reason, are not shown in the figures. The only exception is made for the
one-loop diagram No~7 with a self-contracted solid line in the function
$\langle \varphi \varphi' \rangle$ needed for the calculation of $Z_{5}$,
in which $\tau$ should be retained. Depending of the type of a diagram and
specific way of calculation, the IR regularization is either provided by
the sharp cutoff or is not needed at all (see below). The diagrams with
self-contracted lines $\langle \varphi \varphi' \rangle_{0}$ should also
be discarded according to the general rules of the diagrammatic technique
for the dynamic models of the type (\ref{action}); see \cite{Zinn,Book3}.

Owing to specific properties of our model, many diagrams still shown in the
figures vanish identically or appear UV finite and therefore do not
contribute to the renormalization constants.

Since the velocity field is transverse (divergence-free), the derivative
$\partial_{\parallel}$ at the triple vertex can, if desired, be moved onto
the response field using the integration by parts; see Eq. (\ref{Villon}).
Thus in any diagram involving $N$ {\it external} vertices of the type
$\varphi'(v\partial_{\parallel}) \varphi$, the factor $p\,{}_{\parallel}^{N}$
with external momenta $p_{\parallel}$ will be taken outside the corresponding
integrals over the internal momenta and frequencies. This reduces the
dimension of the integrand and can make the diagram UV finite.
In particular, this makes UV finite the diagrams Nos 1, 3, 4, 13 and 14
(the latter in fact vanishes, see below).

The diagrams Nos 5, 10, 11 and 14 effectively involve closed circuits of
retarded
propagators $\langle \varphi \varphi' \rangle_{0}$ (self-contracted chains
of step functions; see Eq. (\ref{lines3})) and therefore also vanish. Such
effect is well known for dynamic models of the type (\ref{action}) and is
related to causality; see \cite{Zinn,Book3}. However, for the diagrams No
11 and 14 it is crucial here that the propagator of the velocity involves
the $\delta$ function in time. These arguments, however, do not apply to the
one-loop diagram No~6, which requires more accurate consideration
(see below).

There are more diagrams in the functions $\langle\varphi'\varphi'\rangle$
and $\langle\varphi'\varphi\rangle$ having the same topology as the
diagrams Nos 4 and 5, with another placements of the slashes. They also
vanish because of the two reasons explained above, and we do not show them
in the figures.

As a result, in the functions $\langle\varphi'\varphi'\rangle$
and $\langle \varphi' \varphi\varphi\varphi \rangle$ only the diagrams
Nos~2 and~12 are divergent.

Integration over times (or frequencies) in the diagram No~9 leads to the
expression (up to a numerical factor and with implied IR cutoffs)
\begin{eqnarray}
p_{\parallel} \int d\k \int d\q\, \,
\frac{\delta(k_{\parallel})\,q_{\parallel}}
{\left[(\p+\k)^{2}+q^{2}+(\q+\k)^{2}\right] \, q^{2} \,
k_{\bot}^{d-1+\xi}},
\label{NOS}
\end{eqnarray}
where one external momentum $p_{\parallel}$ has already appeared as an
overall factor. Due to the presence of the $\delta$  function in the
integrand, one can replace
$(\p+\k)^{2} \to p_{\parallel}^{2} +(\p_{\bot}+\k_{\bot})^{2} $ in the
denominator. Thus the expansion of the integral in small momenta begins
only with quadratic terms: $p_{\bot}^{2}$ (due to the persisting $SO(d-1)$
invariance) and $p_{\parallel}^{2}$. The total expression (\ref{NOS}) can
only begin with cubic terms and is therefore UV finite, because the possible
superficial divergence of the function $\langle \varphi \varphi' \rangle$
must be quadratic; see section~\ref{sec:Reno}. It remains to note that in the
isotropic case the UV finiteness of such diagram would be guaranteed by the
transversality of the velocity propagator; see the remark in
Ref.~\cite{Ronis}.

Thus we are left with the five UV divergent diagrams Nos~2, 6, 7, 8 and~12.

The analytic expression for diagram No~6 has the form
\begin{eqnarray}
D_{6}(p)= ({\rm i} \sigma p_{\parallel}) ^{2} \,
\int \frac{d\omega}{(2\pi)} \int \frac{d\k}{(2\pi)^{d}} \,
D_{v} (k) \, \frac{1} {  -{\rm i} \sigma \omega + \epsilon(\p-\k)}
\label{D2}
\end{eqnarray}
with $D_{v}$ from (\ref{veloc2}), the prefactor coming from the vertices and
$\epsilon(\k)=\k_{\bot}^{2}+u\k_{\parallel}^{2}$ coming from (\ref{lines3});
the result is independent of the external frequency. Integration over
$\omega$ involves the indeterminacy
\begin{eqnarray}
\int \frac{d\omega}{(2\pi)} \, \frac{1}{-{\rm i} \sigma \omega +
\epsilon(\p-\k)}  = \sigma^{-1}\, \theta(0),
\label{inde}
\end{eqnarray}
where $\theta(0)$ is the step function at the origin. This indeterminacy
reflects the details of the velocity statistics lost in the
white-noise limit (\ref{veloc1}) and it should be carefully resolved;
see e.g. the discussion in the appendix of Ref.~\cite{FGV}. In our case,
the function $\delta(t-t')$ should be understood as the limit of a narrow
function which is necessarily symmetric in $t \leftrightarrow t'$, because
(\ref{veloc1}) is a pair correlation function. Thus the quantity in
(\ref{inde}) must be unambiguously defined by half the sum of the limits:
$\theta(0)=1/2$.
Then after the trivial integration over $\k_{\parallel}$ and using
(\ref{RenD}) one obtains
\begin{eqnarray}
D_{6}(p)= - p\,{}_{\parallel}^{2} \, \frac{wu\mu^{\xi}}{2} \, \int
\frac{d\k_{\bot}}{(2\pi)^{(d-1)}} \,
\frac{1}{k_{\bot}^{(d-1+\xi)}}.
\label{D2a}
\end{eqnarray}
Finally, the integration over $\k_{\bot}$ gives
\begin{eqnarray}
D_{6}(p)= - p\,{}_{\parallel}^{2} \, wu\, (\mu/m)^{\xi} \,
\frac{S_{d-1}}{2(2\pi)^{(d-1)}} \frac{1}{\xi} = -
p\,{}_{\parallel}^{2} \, \frac{wu}{4\pi^2} \,\frac{1}{\xi}+ \ {\rm
UV\ finite\ part},
\label{D2b}
\end{eqnarray}
where $S_d=2\pi^{d/2}/\Gamma(d/2)$ with Euler's $\Gamma$ function is the
surface area of the unit sphere in $d$-dimensional space.

One important remark is in order here. In models with a single UV regulator
(say, model A with $\eps$) the UV singularities manifest themselves as poles
in $\eps$, and the MS scheme is defined such that all the renormalization
constants have the form $Z=1+$ only poles in $\eps$. In models with two
regulators, like $\eps$ and $\xi$ in our case, there are subtleties in
defining the MS scheme: for example, is the ratio $\eps/\xi$ a pole or a
finite quantity. The final physical results must be independent of the choice
of the renormalization scheme. Practical calculations in analogous
two-parameter models (e.g. two-loop calculations for the stochastic
Navier--Stokes equation near two dimensions \cite{local}) confirm that
this is indeed true. In our calculations we always assumed that
$\eps \sim \xi$, treated the combinations like $\eps/\xi$ as UV finite and
did not include them into the renormalization constants; see the last
equality (\ref{D2b}). However, in the leading-order approximations these
subtleties are not too important. In particular, another (and eventually
equivalent) possibility is to include the factor $S_{d-1}/2(2\pi)^{(d-1)}$
(and not only its value at $\eps=0$) into the definition of the new coupling
constant $\tilde w$; see the text below Eq. (\ref{gammaOne}). This will
include all powers of the ratio $\eps/\xi$ from (\ref{D2b}) into the
corresponding renormalization constant $Z_{4}$ without changing the
anomalous dimension $\gamma_{4}$.

The remaining diagrams Nos~2, 7, 8 and~12 do not involve the velocity
propagator $\langle {\bfv} {\bfv} \rangle_{0}$ and can be reduced to the
well-known diagrams of the isotropic model A. Consider the diagram No~8
as an example. The corresponding analytic expression can be represented
in the form
\begin{eqnarray}
\lambda^{2} I(\omega,\, p_{\bot}^{2}+up\,{}_{\parallel}^{2},\,u) =
\lambda^{2}u^{-1} I(\omega,\, p_{\bot}^{2}+up\,{}_{\parallel}^{2},\,u=1).
\label{reskale}
\end{eqnarray}
Here, $\omega$ and $\p=\p_{\bot}+\p_{\parallel}$ are the external frequency
and momentum, $\lambda^{2}$ comes from the vertex factors and $I(\dots)$,
after the integrations over the internal times or frequencies, is represented
as a double integral over the two integration momenta, say, $\k$ and $\q$.
All the momenta enter the integrand via the functions of the type
$\epsilon(\k) = \k_{\bot}^{2}+u\k_{\parallel}^{2}$ coming from the
expressions (\ref{lines3}), (\ref{lines2}) with the replacements $u_{0}\to u$
and $\tau_{0} \to 0$ (see the discussion in the beginning of the appendix).
This explains the fact that $I(\dots)$ depends on $\p$ through the only
scalar argument $p_{\bot}^{2}+up_{\parallel}^{2}$. This also gives the second
equality in (\ref{reskale}) after the rescaling of the integration momenta
$u\k_{\parallel}\to\k_{\parallel}$, $u\q_{\parallel}\to\q_{\parallel}$. The
factor $u^{-1}$ arises from the Jacobians and combines with $\lambda^{2}$ to
give the prefactor $\lambda^{2}u^{-1}=g^{2}$; see Eq.~(\ref{D0}). However,
the quantity on the right-hand side of (\ref{reskale}) is nothing other than
the analytic expression for the diagram No~8 in the isotropic ($u=1$) model
A, while the dependence on $u$ persists in its modified momentum argument
$p^{2} \to p_{\bot}^{2}+up\,{}_{\parallel}^{2}$.

The isotropic integral is represented in the form
\[ I (\omega,\,p,\,u=1) = \left\{ - {\rm i} \sigma \omega {\cal A} +
p^{2} {\cal B} \right\} \, \frac{1}{\eps} +\ {\rm UV\ finite\ part } \]
with known (see e.g. \cite{AV}) dimensionless coefficients ${\cal A}$,
${\cal B}$, which for (\ref{reskale}) gives
\begin{eqnarray}
\lambda^{2} I(\omega,\, p_{\bot}^{2}+up_{\parallel}^{2},\,u) =g^{2} \left\{ -
{\rm i} \sigma\omega {\cal A} + (p_{\bot}^{2}+up\,{}_{\parallel}^{2})
{\cal B}\right\} \, \frac{1}{\eps}  +\ {\rm UV\ finite\ part }.
\label{resk}
\end{eqnarray}
This expression fully determines contribution of the diagram No~8 to the
renormalization constants $Z_{2,3,4}$ in our model (\ref{RenAct}). Similar
considerations determine the contributions of the diagrams 2, 7 and 12 to
the constants $Z_{1,5,6}$ in terms of the known coefficients for the model
A. The latter are well known, but for completeness we will discuss the
corresponding calculational techniques which proved to be useful in the
three-loop calculation in the model A \cite{AV} and might be interesting in
itself.

Two key points are as follows: the convolution of two functions of the form
\begin{equation}
F(\alpha;a) \equiv (-{\rm i}\omega\, a + k^{2})^{-\alpha}
\label{fuNk}
\end{equation}
is a function of the same form,
\begin{equation}
F(\alpha;a) * F(\beta;b) = \theta(ab)\,  K_{2}(\alpha,\beta;a,b)\,
F(\alpha+\beta-d/2-1;a+b)
\label{conv1}
\end{equation}
with the coefficient
\[  K_{2}(\alpha,\beta;a,b) = a^{d/2-\alpha} b^{d/2-\beta}
(a+b)^{\alpha+\beta-d-1} \, \frac{\Gamma(\alpha+\beta-d/2-1)}
{(4\pi)^{d/2} \Gamma(\alpha) \Gamma(\beta)},  \]
while the product of two such functions can be represented as a single
integral of a function of the same form with the aid of the generalized
Feynman formula:
\begin{equation}
F(\alpha;a) \cdot F(\beta;b) = \frac{\Gamma(\alpha+\beta)}
{\Gamma(\alpha)\Gamma(\beta)}  \int_{0}^{1} ds \, s^{\alpha-1}
(1-s)^{\beta-1} F(\alpha+\beta; as+b(1-s)).
\label{conv3}
\end{equation}
Equation (\ref{conv1}) can be obtained from the fact that in the
$\{t,\x\}$ representation the function (\ref{fuNk}) takes on the form
\begin{equation}
F(\alpha;a) \to \frac{\theta(t\,{\rm sign} (a))\, a^{d/2-\alpha}}
{(4\pi)^{d/2}\, \Gamma(\alpha)}\, t^{\alpha-d/2-1} \,
\exp \left\{ - \frac{ax^{2}}{4t} \right\},
\label{xt}
\end{equation}
and the product of such functions (which corresponds to the convolution
of their Fourier transforms) is obviously a function of the same form.
Note that for $ab<0$ the convolution (\ref{conv1}) vanishes because it
corresponds to the product of a retarded and an advanced functions of
the form (\ref{xt}). In the same manner, for the convolution of three
functions (\ref{fuNk}) one obtains
\begin{equation}
F(\alpha;a) * F(\beta;b) * F(\gamma;c) = K_{3}\, (\alpha,\beta,\gamma;a,b,c)
F(\alpha+\beta+\gamma-d-2)
\label{conv4}
\end{equation}
with the coefficient
\[ K_{3} (\alpha,\beta,\gamma;a,b,c) = \theta(ab) \theta(bc)\,
\frac{\Gamma(\alpha+\beta+\gamma-d-2)}{(4\pi)^{d/2}
\Gamma(\alpha) \Gamma(\beta) \Gamma(\gamma)} \times \]
\[ \times a^{d/2-\alpha} b^{d/2-\beta} c^{d/2-\gamma}
(a+b+c)^{\alpha+\beta+\gamma-3d/2-2}, \]
and so on.

In the notation (\ref{fuNk}), the bare propagators of the scalar fields
$\varphi$, $\varphi'$  can be written as
\[ \langle \varphi\varphi' \rangle_{0} = F(1,\sigma), \quad
\langle \varphi'\varphi \rangle_{0} = F(1,-\sigma), \quad
\langle \varphi\varphi \rangle_{0} = \int^{\sigma}_{-\sigma} du\, F(2;u). \]
In the last expression we used the equation (\ref{conv3}) with
$a=-b=\sigma$, $\alpha=\beta=1$ and introduced the new integration
variable as $u = \sigma(1-2s)$.

Consider again the diagram No~8 as an example of the calculation of the pole
part in $\eps$. It is convenient to set $\sigma=1$ (the dependence on
$\sigma$ can easily be restored by dimensionality) and discard the factors
like $(4\pi)^{-d/2}$ which always combine with $g$ to form the new coupling
constant $\tilde g$; see the text below Eq. (\ref{gammaOne}). Then the
analytic expression for the diagram No~8 reads (we omit the factor $g^{2}$):
\begin{equation}
D_{8} = \int^{1}_{-1} du_{1} \int^{1}_{-1} du_{2} \, F(2,u_{1}) *
F(2,u_{2}) * F(1,1).
\label{D9a}
\end{equation}
Applying the reference formula (\ref{conv4}) to the integrand of (\ref{D9a})
gives
\begin{eqnarray}
D_{8} &=& \Gamma(3-d) \int^{1}_{0} du_{1} \int^{1}_{0} du_{2} \,
u_{1}^{2-d/2} u_{2}^{2-d/2} \times
\nonumber \\
&\times&  (u_{1}+u_{2}+1)^{3-3d/2} \, F(3-d,u_{1}+u_{2}+1).
\label{D9b}
\end{eqnarray}
Note that only positive values of $u_{1,2}$ survive in (\ref{D9b}) due to
the $\theta$ functions in (\ref{conv4}).

In order to extract the pole part of (\ref{D9b}), it is sufficient
to replace $\Gamma(3-d)=\Gamma(-1+\eps) \to -1/\eps$ and to set
$\eps=0$ in the integrand; this gives:
\begin{equation}
D_{8} = - \frac{1}{2\eps} \left[ -{\rm i} \omega I_{2}
+ p^{2} I_{3} \right] + \ {\rm UV\ finite\ part},
\label{kuku}
\end{equation}
with the coefficients
\begin{equation}
I_{n} = \int^{1}_{0} du_{1} \int^{1}_{0} du_{2} \, \frac{1}
{(u_{1}+u_{2}+1)^{n}} = \int^{2}_{0} du\, p(u) \, \frac{1}{(1+u)^{n}}
\label{Jn}
\end{equation}
and the weight function
\[ p(u)= \cases{ u  &  for $u<1$, \cr 2-u  &  for $u>1$. \cr  } \]
The integrals in (\ref{Jn}) are easily calculated directly, but it is
instructive to use the following trick, which proved extremely useful in
the higher-order calculations. Let us introduce the generating function
\begin{equation}
I(z) = \int^{2}_{0} du\, p(u) \, \frac{1}{z+u}\,.
\label{Jz}
\end{equation}
It is easily seen that the integrals $I_{2}$ and $I_{3}$ from (\ref{Jn}) are
the second and the third coefficients in the expansion of the quantity
$I(1-\delta)$ in $\delta$. Calculation of the integral (\ref{Jz}) up to
irrelevant constant terms gives
\[ I(z) \simeq z \ln\,\frac{z(z+2)}{(z+1)^{2}} +2 \ln\, \frac{z+2}{z+1}, \]
and the expansion
\[ I(1-\delta) = \const + \delta \ln(4/3) + \delta^{2} /6 + O(\delta^{3})\]
gives the desired coefficients $I_{2,3}$ in (\ref{kuku}), which in their
turn determine the coefficients ${\cal A}$, ${\cal B}$ in (\ref{resk}).

Once a renormalization constant $Z_{i}$ has been calculated, the
corresponding anomalous dimension is readily found from the
relation
\begin{equation}
\gamma_{i} = (\beta_{g}\partial_{g}+\beta_{w}\partial_{w}) \ln Z_{i}
= - (\eps\D_{g}+\xi\D_{w}) \ln Z_{i}.
\label{GfZ}
\end{equation}
In the first equality, we used the definition (\ref{RGF1}), expression
(\ref{RG2}) for the operation $\Dm$ in renormalized variables, and the
fact that the $Z$'s depend only on the two completely dimensionless coupling
constants $g$ and $w$. In the second equality, we retained only the
leading-order terms in the $\beta$ functions (\ref{betagw}), which is
sufficient for our approximation. The factors $\eps$ and $\xi$ in
(\ref{GfZ}) cancel the corresponding poles contained in $\ln Z_{i}$,
which leads to the final UV finite expressions for the anomalous dimensions,
given in (\ref{gammaOne}) and (\ref{gammaTwo}).

\end{document}